\documentclass[12pt,english,preprint,superscriptaddress,showpacs,nofootinbib]{revtex4-1}
\usepackage[T1]{fontenc}
\usepackage[utf8x]{inputenc}
\usepackage{amsmath}
\usepackage{graphicx}
\usepackage{amssymb}
 \usepackage{amsfonts}
 \usepackage{epsfig}
 \usepackage{subfigure}
 \usepackage{graphicx}
 \usepackage{pst-grad} 
 \usepackage{pst-plot} 
 \usepackage{caption}
 \usepackage{mathrsfs} 
 \usepackage{float}

\makeatletter
\@ifundefined{textcolor}{}
{%
 \definecolor{BLACK}{gray}{0}
 \definecolor{WHITE}{gray}{1}
 \definecolor{RED}{rgb}{1,0,0}
 \definecolor{GREEN}{rgb}{0,1,0}
 \definecolor{BLUE}{rgb}{0,0,1}
 \definecolor{CYAN}{cmyk}{1,0,0,0}
 \definecolor{MAGENTA}{cmyk}{0,1,0,0}
 \definecolor{YELLOW}{cmyk}{0,0,1,0}
 }

\makeatother

\usepackage{babel}

\begin{document}

\title {The Multi-Layer Multi-Configuration Time-Dependent Hartree Method for Bosons: Theory, Implementation and Applications}
\thanks{L. Cao and S. Kr\"{o}nke have contributed equally to this work.}
\author{Lushuai Cao}
\email{lcao@physnet.uni-hamburg.de}
\affiliation{Zentrum f\"{u}r Optische Quantentechnologien, Universit\"{a}t Hamburg, Luruper Chaussee 149, D-22761 Hamburg, Germany}
\affiliation{The Hamburg Centre for Ultrafast Imaging,Luruper Chaussee 149,D-22761 Hamburg, Germany}
\author{Sven Kr\"{o}nke}
\email{Sven.Kroenke@physnet.uni-hamburg.de}
\affiliation{Zentrum f\"{u}r Optische Quantentechnologien, Universit\"{a}t Hamburg, Luruper Chaussee 149, D-22761 Hamburg, Germany}
\author{Oriol Vendrell}
\email{oriol.vendrell@cfel.de}
\affiliation{Center for Free-Electron Laser Science, DESY, Notkestrasse 85, D-22607 Hamburg, Germany}
\affiliation{The Hamburg Centre for Ultrafast Imaging,Luruper Chaussee 149,D-22761 Hamburg, Germany}
\author{Peter Schmelcher}
\email{pschmelc@physnet.uni-hamburg.de}
\affiliation{Zentrum f\"{u}r Optische Quantentechnologien, Universit\"{a}t Hamburg, Luruper Chaussee 149, D-22761 Hamburg, Germany}
\affiliation{The Hamburg Centre for Ultrafast Imaging,Luruper Chaussee 149,D-22761 Hamburg, Germany}

\begin{abstract}
We develop the multi-layer multi-configuration 
time-dependent Hartree method for bosons (ML-MCTDHB), a variational numerically exact ab-initio method for studying the quantum dynamics and stationary properties of general bosonic systems. ML-MCTDHB takes advantage of the permutation symmetry of identical bosons, which allows for investigations of the quantum dynamics from few to many-body systems. Moreover, the multi-layer feature enables ML-MCTDHB to describe mixed bosonic systems consisting of arbitrary many species. Multi-dimensional as well as mixed-dimensional systems can be accurately and efficiently simulated via the multi-layer expansion scheme. We provide a detailed account of the underlying theory and the corresponding implementation. We also demonstrate the superior performance by applying the method to the tunneling dynamics of bosonic ensembles in a one-dimensional double well potential, where a single-species bosonic ensemble of various correlation strengths and a weakly interacting two-species bosonic ensemble are considered.
\end{abstract}


\maketitle

\section{Introduction}

The numerical simulation of the quantum dynamics of strongly correlated many-body systems is a topic of widespread and pronounced interest. 
Such simulations represent a tough task in general due to the exponential scaling of the state space: taking an
interacting $N$-particle system for example, and allowing each particle to occupy $M$ single particle states, the
many-body Hilbert space turns out to be $M^N$ dimensional for a system of
distinguishable particles. When going to higher particle numbers, one, hence,
has to truncate the Hilbert space in one or another way. Moreover, this scaling problem has 
serious consequences for the simulation of the dynamics: In the course of time, the
system state will move through different subspaces of the total Hilbert space
in general. If one now tries to expand the systems state with respect to a
time-independent basis, many basis states will be needed in order to span the
relevant subspace for the whole propagation time.
So for reducing the number of
basis vectors, one should better employ a time-dependent, moving basis, in
which the instantaneous system state can be optimally represented. 

In the multi-configuration time-dependent Hartree method (MCTDH), such
a co-moving basis is provided by time-dependent Hartree products
\cite{MMC90,BJWM00}. The equations of motion for the single particle functions
(SPFs) building up the
time-dependent Hartree products as well as for the expansion coefficients are
obtained by means of the Dirac-Frenkel variation principle, which ensures a
variationally optimal representation of the total wave function at any instant. 
Using Hartree products as the many-body basis states, MCTDH is well suited for
dealing with distinguishable particles. Originally developed in the context of
 quantum molecular dynamics, MCTDH has successfully been applied
to  an enormous diversity of problems concerning the vibrational and wave
packet dynamics of molecules (cf. \cite{HDMey09,HDMey12} and
reference therein). Moreover, MCTDH
turned out to be also
powerful for simulating strongly correlated few-body systems of ultracold,
indistinguishable bosonic atoms in traps, see e.g. \cite{ZMS08, CBZS11}.

MCTDH has been extended for dealing with higher particle numbers in several
ways: In multi-layer MCTDH (ML-MCTDH), degrees of freedom are combined to
higher-dimensional ones. The MCTDH expansion scheme is then applied to theses
combined modes and their constituting degrees of freedom in a cascade
\cite{WT03, Manth08}. This
scheme is particularly fruitful for scenarios where there are strong
correlations within certain subsystems while the inter-subsystem correlations
are relatively weak \cite{VHDM11}. Especially, many system-bath problems are of
this type. In
such a situation, the necessary number of time-dependent basis functions can be
reduced by combining the strongly correlated degrees of freedom in an ML-MCTDH
expansion. As MCTDH, ML-MCTDH is based on expansions in terms of Hartree
products and, thus, ignores possible particle exchange symmetries of the
system. When considering an interacting bosonic bath, the resulting redundancies
in the configuration space limit the feasible numbers of the indistinguishable
degrees of freedom and SPFs.

In contrast to this, the MCDTH method for fermions (MCTDHF) and for bosons
(MCTDHB) reduce the number of expansion coefficients by expanding the total
wave function in terms of slater determinants and permanents
\cite{mctdhf_1,ASC07,ASC08,SAC10}, respectively.
MCTDHF has been applied in the context of quantum chemistry as well as for
molecules
in strong laser fields \cite{mctdhf_2,mctdhf_3,mctdhf_4,mctdhf_5}. The
crossover from the Gross-Pitaevskii to
the
many-body regime has been studied for several ultracold bosonic systems by
means of MCTDHB (see ref. \cite{SSAC09, SAC11} and the refs. therein). 
There is even a formalism, which
unifies MCTDHB and MCTDHF for
dealing, $e.g.$ with bose-bose, bose-fermi and fermi-fermi mixtures \cite{MCTDHBB,ASSC11}.
Furthermore, an extension for considering particle conversion, $i.e.$ chemical reactions, in a multi-species
system has been developed \cite{ML_partconv}.

ML-MCTDH features a high flexibility with respect to the wave function
expansion: Any way of combining degrees of freedom to higher dimensional ones
in a cascade is allowed and leads to a specific wave function ansatz. As, in
the case of indistinguishable particles, many of these mode-combination schemes
do not reflect the indistinguishability of the particles, the particle exchange
\mbox{(anti-)} symmetry can, in general, not be incorporated in ML-MCTDH, making
ML-MCTDH incompatible with MCTDHB and MCTDHF. There is, however, an exception:
In \cite{WT09}, a second quantization representation of ML-MCDTH (ML-MCTDH-SQR)
has been derived, which has been applied to charge transport
problems \cite{ML-SQR1,ML-SQR2}, for example. The ML-MCTDH-SQR method is based
on the factorization of the Fock space for $m$ single particle modes into $m$
sub-Fock spaces corresponding to just one single particle mode each, which
leads to the product structure required by ML-MCTDH.
By the virtue of this product structure, the $m$ modes can be identified with the
"lower dimensional" DOF of a ML-MCTDH setting and any mode combination of these DOF
to higher dimensional DOF in a cascade becomes possible.
In this way, ML-MCTDH-SQR benefits
from the flexible cascade scheme of ML-MCTDH while keeping the SPF, constituting
the Fock states, time-independent.

In this work, an alternative symbiosis employing the bosonic symmetry and a multi-layer ansatz is presented,
in which the SPFs of the bosons are time-dependent. The price one has to pay
for this is that only two expansion schemes (and any
combination thereof) are possible: (i) a multi-layer expansion with respect to
the different bosonic species for optimally taking account inter- and
intra-species
correlations and (ii) a particle multi-layer expansion in which the single
particle functions are further expanded in the MCTDH form for efficiently
simulating bosons in two- and three-dimensional traps. This method, termed
multi-layer MCTDHB (ML-MCTDHB), manifests itself as a numerically exact ab-initio
tool for the quantum dynamics as well as stationary properties of bosonic ensembles
consisting of arbitrary many species in various dimensions. This paper is devoted to
the detailed derivation of the underlying theory, the description of the implementation,
as well as the account of algorithmic features of the method. While Ref \cite{mlb_3spc}
contains a brief account of the theory of ML-MCTDHB and focuses on its applications to
the nonequilibrium quantum dynamics of mixtures of ultracold atoms, the present work
provides the complete theoretical framework necessary for the reader to thoroughly
understand and apply the theory. Furthermore we provide relevant technical aspects,
address the strategy of an efficient implementation and demonstrate the efficiency of ML-MCTDHB
by comparing the number of expansion coefficients with
the existing methods such as MCTDHB for bose-bose mixtures by applying it to some relevant 
example systems. 

This paper is organized as follows: in section \ref{sec_theo}, we introduce two different schemes of 
ML-MCTDHB, the species ML-MCTDHB (in section \ref{subsec_species}) and the particle ML-MCTDHB 
(in section \ref{subsec_par}), which
are for bosonic mixtures and bosonic systems in high dimensions, respectively. The combination 
of these two schemes leading to the general ML-MCTDHB theory is then
introduced in \ref{subsec_intg}. In section \ref{subsec_scaling} and \ref{subsec_symm}, we
comment on the scaling property and the symmetry preservation of ML-MCTDHB, respectively.
In section \ref{sec_impl} we discuss some details of the implementation of the method,
including the implementation of the second quantization scheme (\ref{subsec_ns})
and the applicabilities of the method (\ref{subsec_app}). In section
\ref{sec_exampl}, the application of ML-MCDTHB is illustrated by two examples.
We conclude
with a summary, discussing also an extension of ML-MCTDHB to fermionic species
in section \ref{sec_concl}.
\section{Theory}\label{sec_theo}
\subsection{ML-MCTDHB for Bosonic Mixtures}\label{subsec_species}
\subsubsection{Hamiltonian and Ansatz}

 ML-MCTDHB is designed for the most general bosonic mixtures containing an arbitrary number of bosonic
 species, with all possible types of single particle potentials for each species as well as intra-/inter-species
 interaction potentials. The bosonic species can refer to different bosonic quasi-particles, chemical species,
 isotopes or atoms of the same element prepared in different hyperfine bosonic states, which have been widely
 realized in experiments \cite{mix_KRb,mix_spin1,mix_spin2}. ML-MCTDHB can treat bosons with various degrees of
 freedom (DOF), including one- to three-dimensional spatial DOF as well as spatial plus internal (e.g. the spinor) DOF.
 The Hamiltonian of such bosonic mixtures can be expressed in the general form of 
 \begin{equation}\label{Ham_1st}
  \hat H = \sum_\kappa \sum_i \hat H^0_\kappa(i)+\sum_\kappa\sum_{i<j}\hat V_\kappa(i,j)+\sum_{\kappa_1<\kappa_2}\sum_{i_{\kappa_1},j_{\kappa_2}}\hat W_{\kappa_1,\kappa_2}(i_{\kappa_1},j_{\kappa_2}).
 \end{equation}
  The first term of the Hamiltonian contains the single particle Hamiltonian for all the bosons
  in each species with $\hat H^0_\kappa=\hat T_\kappa+\hat U_\kappa$, 
  and $\hat T_\kappa$ as well as $\hat U_\kappa$ refer to the kinetic energy and single particle
  potentials of the $\kappa$ species bosons, respectively. Particularly $\hat U_\kappa$ can be a spatial
  potential and also a spin-orbit coupling potential for bosons with spin DOF. The second and third terms
  of equation (\ref{Ham_1st}) are the intra- and inter-species interactions. The intra-/inter-species
  interactions, $\hat V_\kappa$ and $\hat W_{\kappa_1,\kappa_2}$ respectively, can be of arbitrary form
  such as the contact interaction, the dipolar interaction or the Coulomb interaction, which are widely
  studied in various physics and chemistry fields. In this work, the interactions are restricted to the
  two-body interactions. Moreover, since ML-MCTDHB is a variational ab-initio method for the time-dependent
  Schr\"{o}dinger equation, all the potentials can be time-dependent in ML-MCTDHB.
    
  ML-MCTDHB applies a multi-layer ansatz for the bosonic mixture of arbitrary many species. To introduce the
  multi-layer ansatz, we consider a general bosonic mixture containing S species, and bosons of each species
  evolve in a fixed single particle Hilbert space, which is spanned by a set of time-independent basis
  functions, $e.g.$, $\{|r^\kappa_j\rangle\}_{j=1}^{\mathscr{M}_\kappa}$ for the bosons of species $\kappa$, which
  we call $\kappa$ bosons for brevity in the following. The Hilbert space of the total system is chosen as the
  direct product of a given Hilbert space of each species, as $\mathscr{H}=\prod_{\kappa=1}^S\mathscr{H}^\kappa$,
  where $\mathscr{H}$ and $\mathscr{H}^\kappa$ refer to the Hilbert space of the total system and the $\kappa$ species,
  respectively. Consequently, the total wave vector $|\Psi\rangle$ of the system, temporally evolving in $\mathscr{H}$,
  is expanded in Hartree products of the vectors in each $\mathscr{H}^\kappa$, as 
   \begin{equation}\label{mlb_ansatz}
    |\Psi\rangle = \sum_{i_1=1}^{M_1} ... \sum_{i_S=1}^{M_S}
      A^1_{i_1,...,i_S}
      \prod_{\kappa=1}^S|\phi^{2;\kappa}_{i_\kappa}\rangle.
  \end{equation}
  $|\phi^{2;\kappa}_{i_\kappa}\rangle$ refers to a vector in $\mathscr{H}^\kappa$,
  and $A^1_{i_1,...,i_S}$ is the time-dependent expansion coefficient. For simplicity
  we omit all time-dependencies in the notation in this paper.
  
  In ML-MCTDHB, $\{|\phi^{2;\kappa}_{i}\rangle\}|_{i=1}^{M_\kappa}$ for all species,
  $i.e.$ $\kappa\in[1,S]$, are set to be time-dependent, and they are evolving within
  the corresponding Hilbert space $\mathscr{H}^\kappa$. Instead of taking the basis vectors
  of $\mathscr{H}^\kappa$ as the direct product of $|r^\kappa\rangle$ of all the $\kappa$ bosons,
  a set of time-dependent SPFs $\{|\phi^{3;\kappa}_i\rangle\}|_{i=1}^{m_\kappa}$ is assigned to
  each of the $\kappa$ bosons, and the basis vectors of $\mathscr{H}^\kappa$ are then chosen as
  the permanent states with respect to $\{|\phi^{3;\kappa}_i\rangle\}|_{i=1}^{m_\kappa}$ in the second quantization picture, as 
  \begin{equation}\label{permanent}
    |\vec n_\kappa\rangle = (N_\kappa!n_{1}!...n_{m_\kappa}!)^{-\frac{1}{2}}\sum_{\pi\in S(N_\kappa)} 
    |\phi^{3;\kappa}_{i_{\pi(1)}}\rangle_1 \cdots |\phi^{3;\kappa}_{i_{\pi(N_\kappa)}}\rangle_{N_\kappa}.
  \end{equation}
  $|\vec{n}_\kappa\rangle=|(n_1,...,n_{m_\kappa})\rangle$ is the permanent state, with $n_i$ the
  boson number in the state $|\phi^{3;\kappa}_{i}\rangle$, and $|\phi^{3;\kappa}_{i}\rangle_j$ denotes
  that the $j$-th $\kappa$ boson is in the state $|\phi^{3;\kappa}_{i}\rangle$. The summation runs over
  all permutations $\pi$ of the first $N_\kappa$ integers, i.e. $\pi:[1,N_\kappa]\rightarrow [1,N_\kappa]$.
  $\mathscr{H}^\kappa$ is then spanned by the basis vectors of $\{|\vec n_\kappa\rangle\}$,
  and $|\phi^{2;\kappa}_{i_\kappa}\rangle$ is expanded as 
   \begin{equation}\label{spc_spf}
    |\phi^{2;\kappa}_{i}\rangle = \sum_{\vec n_\kappa|N_\kappa}
	  A^{2;\kappa}_{i;\vec n_\kappa} \;|\vec n_\kappa\rangle.
  \end{equation}
  The notation $\vec n_\kappa|N_\kappa$ indicates the constraint that the occupation
  numbers $n_i$ have to sum up to the number of $\kappa$ bosons, i.e. $N_\kappa$.
  
  The single-particle functions (SPFs), $\{|\phi^{3;\kappa}_{i}\rangle\}|_{i=1}^{m_\kappa}$,
  are in turn evolving in the single particle Hilbert space spanned by the time-independent
  basis $\{|r^\kappa_j\rangle\}_{j=1}^{\mathscr{M}_\kappa}$, as 
    \begin{equation}\label{par_spf}
   |\phi^{3;\kappa}_i\rangle=\sum_{j=1}^{\mathscr{M}_\kappa}A^{3;\kappa}_{i;j}|r^\kappa_j\rangle.
  \end{equation}
  Please note that the expansion scheme (\ref{mlb_ansatz}-\ref{par_spf}) is based on a cascade
  of truncations. Having the number of time-independent basis states $\mathcal{M}_\kappa$ fixed,
  one may consider any number of single particle functions $m_\kappa =1,...,\mathcal{M}_\kappa$ and
  any number of species states $M_\kappa=1,..., \binom{N_\kappa+m_\kappa-1}{m_\kappa-1}$,
  so that $m_\kappa$ and $M_\kappa$ serve as numerical control parameters for simulating bosonic
  ensembles of different intra- and inter-species correlations.
 
  The combination of equations (\ref{mlb_ansatz}-\ref{par_spf}) complete the cascade expansion
  of the system wave vector $|\Psi\rangle$ with respect to the fixed basis $\{|r_\kappa\rangle\}$, and
  a concept of multi-layer structure can be identified: $|\Psi\rangle$ corresponds to the top layer
  and $\{|\phi^{2;\kappa}_i\rangle\}$, $\{|\phi^{3;\kappa}_i\rangle\}$ form the species and particle
  layer, respectively.
  $\{|r_\kappa\rangle\}$ are related to the physical or in accordance with the MCTDH
  terminology \cite{BJWM00} primitive DOF, $e.g.$ the spatial and internal DOF,
  while $\{|\phi^{2;\kappa}_i\rangle\}$ and $\{|\phi^{3;\kappa}_i\rangle\}$ are related to the
  so-called logical DOF on the species and particle layers, respectively. Each logical DOF on
  the species layer represents one bosonic species, and each logical DOF on the particle layer
  represents a boson of the corresponding species. In the following $\{|r_\kappa\rangle\}$ are
  termed as the primitive basis, with $\{|\phi^{3;\kappa}_i\rangle\}$ 
  and $\{|\phi^{2;\kappa}_i\rangle\}$ named as the particle and 
  species SPFs of corresponding logical DOF of related layers. The class of wave function
  decomposition is denoted as the \textit{species multi-layer MCTDHB} ansatz.
  
  In practice, $|r^\kappa\rangle$ can be taken as the state of one boson locating at
  position $r^\kappa$ in space, and this choice of the primitive basis vectors $|r^\kappa\rangle$ can
  naturally cover high-dimensional systems. Consider, for instance, the $\kappa$ bosons possessing
  three-dimensional spatial DOF plus one spin DOF, $i.e.$, $r^\kappa=(x^\kappa,y^\kappa,z^\kappa,s^\kappa)$,
  then the vector $|r^\kappa\rangle$ is seen as the product
  $|x^\kappa\rangle|y^\kappa\rangle|z^\kappa\rangle|s^\kappa\rangle$, denoting one boson located at
  spatial position $(x^\kappa,y^\kappa,z^\kappa)$ and in spin state $s^\kappa$. We refer to such a
  treatment of higher dimensional SPFs $|\phi^{3;\kappa}_i\rangle$ by means of a product basis of
  one-dimensional time-independent basis states as primitive mode combination. A more efficient treatment
  of high-dimensional systems with ML-MCTDHB will be introduced in the following section.
  
\begin{figure}
\includegraphics[width=0.6\textwidth]{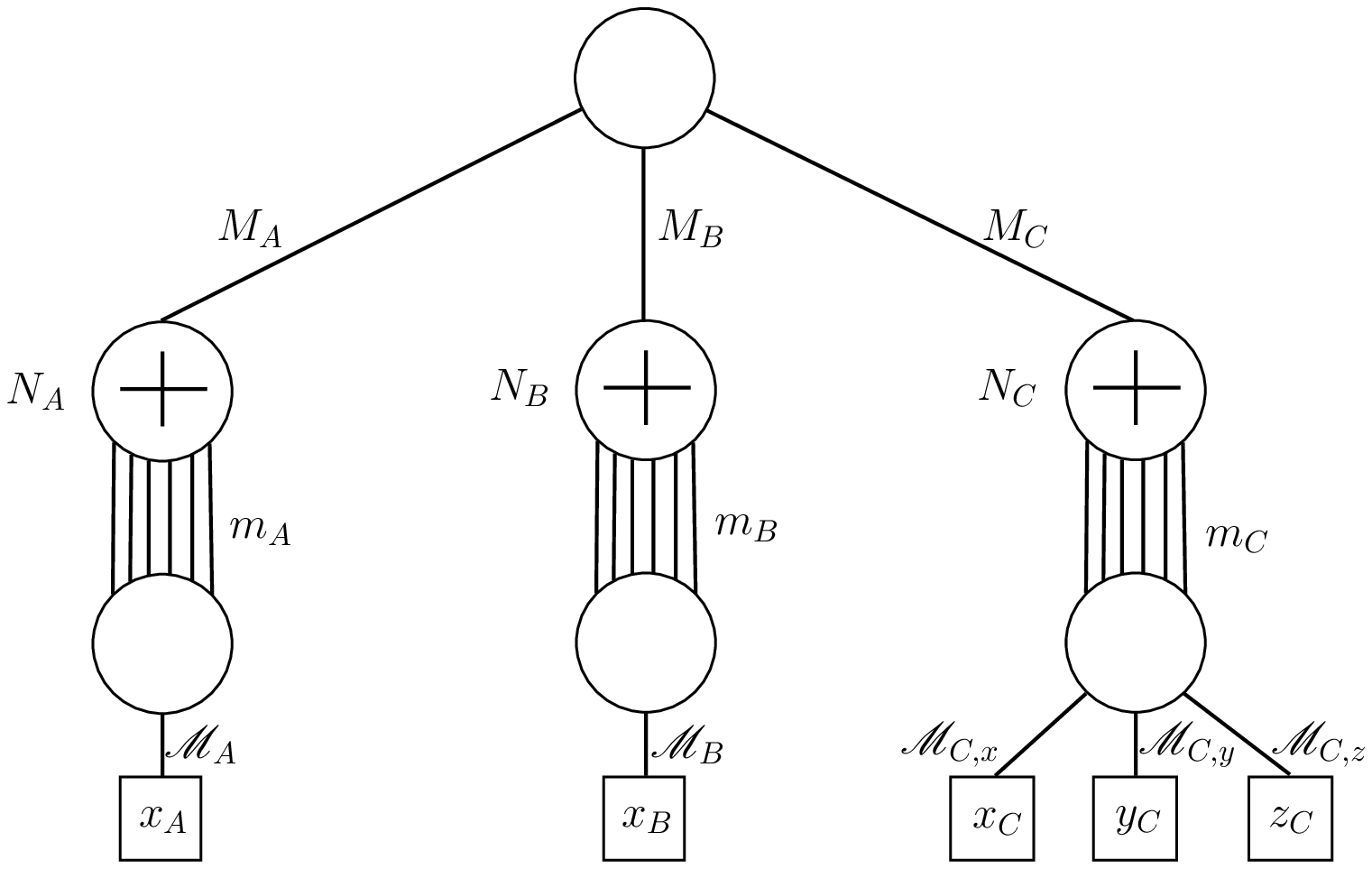}
\caption{The tree diagram of a three-species bosonic system. The three species are labeled as A, B and C bosons,
with primitive DOF of $x_A$, $x_B$ and $(x_C,y_C,z_C)$, respectively. Corresponding to the ML-MCTDHB ansatz,
a three-layer tree diagram is shown, containing the top layer, the species layer, the particle layer
from top to bottom. The nodes on each non-top layer correspond to different logical DOF, and the primitive DOF are
given by the square nodes at the bottom. The "+" inside the species nodes denotes the species to be a bosonic species.}
\label{tree_sketch}
\end{figure}

    Let us summarize the species ML-MCTDHB with the introduction of the tree diagram notation,
  in analogy to \cite{Manth08}. An example of the tree structure is shown in Figure \ref{tree_sketch},
  for which we consider a three-species bosonic mixture containing the bosonic species A, B and C.
  The primitive basis of A and B bosons is spanned with respect to a single spatial DOF
  $\{x^A\}$ and $\{x^B\}$, respectively, with the DOF of the y- and z-dimension frozen out by strong
  confinement potentials,  while the C bosons are in three-dimensional space denoted by $(x^C,y^C,z^C)$.
  The tree diagram is composed of various nodes and links connecting the nodes. The nodes in the tree refer
  to the primitive as well as logical DOF, and the links indicate the expansion relation between corresponding
  DOF. In figure \ref{tree_sketch}, the top node refers to the state vector of the whole system $|\Psi\rangle$,
  and it forms the top layer of the tree. The second layer is the species layer formed by the species nodes,
  and each species node corresponds to the logical DOF of one bosonic species associated with the species
  SPFs $\{|\phi^{2;\kappa}_i\rangle\}$ ($\kappa=A,B,C$). One layer below, $i.e.$ the third layer, is the
  particle layer, of which each node refers to a particle logical DOF with the particle SPFs
  $\{|\phi^{3;\kappa}_i\rangle\}$. The nodes below the particle nodes relate to the primitive spatial DOF,
  which are indicated inside boxes. It now becomes clear that a circle node refers to a logical DOF associated with
  time-dependent basis vectors, and a square refers to a primitive DOF associated with time-independent basis vectors.
  The links between nodes in adjacent layers refer to the expansions
  given in equations (\ref{mlb_ansatz}-\ref{par_spf}), for instance, the three branches leading out of the
  top node correspond to the expansion of $|\Psi\rangle$ by equation (\ref{mlb_ansatz}).
  Particularly,
  the multiple vertical lines linking the species and particle nodes indicate that the species node represents
  a logical DOF to which many bosons rather than a single one are combined.
  The number given along each link denotes
  the number of SPFs or primitive vectors of the node at the bottom side of the link.
  Moreover, the '+' sign in the species node indicates the related
  node as a logical DOF of a bosonic species, and the number on the
  left side of each species node indicates the number of bosons of this species.
  In this way the tree diagram offers the complete information of the ML-MCTDHB ansatz,
  in a visualized manner, and the superscripts of the SPFs and expansion coefficients,
  $|\phi^{\alpha;\kappa}_i\rangle$ and $A^{\alpha;\kappa}_{i;I}$ indicate that the corresponding DOF is
  the $\kappa$-th node on the $\alpha$-th layer.

\subsubsection{Derivation and General Form of the Equations of Motion for the Expansion Coefficients}\label{ssubsec_eq}

  In this section we derive the equations of motion for the coefficients of the ML-MCTDHB ansatz
  (\ref{mlb_ansatz}-\ref{par_spf}) for a general mixture containing $S$ bosonic species, of which
  the Hamiltonian is given by equation (\ref{Ham_1st}) and the time-independent primitive basis of
  each species is chosen as $\{|r^\kappa_j\rangle\}_{j=1}^{\mathscr{M}_\kappa}$.

  The time propagation of the system (both in real and imaginary time) is given by the time evolution
  of all the coefficients in the ansatz, which can be derived from the Dirac-Frenkel variational
  principle \cite{Dirac,Frenkel,McLachlan} 
  \begin{equation} \label{df_principle}
   \langle \delta \Psi|(i\partial_t-\hat H)|\Psi\rangle=0.
  \end{equation}
  Here we adapt the natural units and set $\hbar=1$. The Dirac-Frenkel variational principle turns out to be equivalent to the
Lagrangian and to the McLachlan's principle \cite{McLachlan,VarPrinc,DF_equiv}. The latter
essentially says: Propagate your ansatz $|\Psi\rangle$ according to
 $i\partial_t|\Psi\rangle =  |\Theta\rangle$
with $|\Theta\rangle$ minimizing $|||\Theta\rangle-\hat H|\Psi\rangle||^2$. This ensures that we
obtain a variationally optimal wave function within our class of ansatzes.

To calculate $|\delta\Psi\rangle$ and $i\partial_t|\Psi\rangle$ in (\ref{df_principle}), we firstly introduce a new expression of the permanent state as 
\begin{equation}
 |\vec{n}_\kappa\rangle=\sum_i\sqrt{\frac{n_i}{N_\kappa}}|\phi^{3;\kappa}_i\rangle_1|\vec{n}_\kappa-\hat{i}\rangle_{\overline{1}},
\end{equation}
 where $|\phi^{3;\kappa}_i\rangle_1$ refers to the first boson of the $\kappa$ species in the state of
 $|\phi^{3;\kappa}_i\rangle$ and $|\vec{n}_\kappa-\hat{i}\rangle_{\overline{1}}$ is the permanent state of
 the remaining $N_\kappa-1$ bosons of the $\kappa$ species, with
 $|\vec{n}_\kappa-\hat{i}\rangle_{\overline{1}}=|(n_1,...,n_i-1,...,n_{m_\kappa})\rangle_{\overline{1}}$.
 This is equivalent to dividing the Hilbert space of the $\kappa$ species as
 $\mathscr H^\kappa=\mathscr H^{\kappa;1}\otimes\mathscr H^{\kappa;\overline{1}}$,
 where $\mathscr H^{\kappa;1}$ is the Hilbert space of the first $\kappa$ boson spanned by the vectors of
 $\{|\phi^{3;\kappa}_i\rangle_1\}|_{i=1}^{m_\kappa}$ and $\mathscr H^{\kappa;\overline{1}}$ is the Hilbert
 space of the remaining $N_\kappa-1$ bosons, spanned by the vectors of
 $\{|\vec n_\kappa\rangle_{\overline{1}}\equiv|(n_1,...,n_{m_\kappa})\rangle_{\overline{1}}|\sum_{i=1}^m n_i=N_\kappa-1\}$.
 Similarly we can define the permanent state
 $|\vec n_\kappa\rangle_{\overline{l}}$, which includes all the $\kappa$ bosons except the $l$-th $\kappa$ boson.

Now $|\delta\Psi\rangle$ and $\partial_t|\Psi\rangle$ can be written as 
  \begin{align}\label{delta}
   |\delta\Psi\rangle = \sum_{I}\delta A^1_{I}|I\rangle+\sum_{\kappa,i,\vec{n}_\kappa}\delta A^{2;\kappa}_{i;\vec{n}_\kappa}|\vec{n}_\kappa\rangle|\tilde{\Psi}^{2;\kappa}_i\rangle
    +\sum_{\kappa,l,i,s}\delta A^{3;\kappa}_{i;s}|r^\kappa_s\rangle_l|\tilde{\Psi}^{3;\kappa}_i\rangle_{\overline{l}};\\\label{partial}
   \partial_t|\Psi\rangle = \sum_{I}(\partial_t A^1_{I})|I\rangle+\sum_{\kappa,i,\vec{n}_\kappa}(\partial_t A^{2;\kappa}_{i;\vec{n}_\kappa})|\vec{n}_\kappa\rangle|\tilde{\Psi}^{2;\kappa}_i\rangle
    +\sum_{\kappa,l,i,s}(\partial_t A^{3;\kappa}_{i;s})|r^\kappa_s\rangle_l|\tilde{\Psi}^{3;\kappa}_i\rangle_{\overline{l}}.   
  \end{align}
  In equations (\ref{delta}) the first term refers to the expansion of $|\delta\Psi\rangle$ with respect
  to the top-node coefficients with $|I\rangle\equiv\prod^{s}_{\kappa=1}|\phi^{2;\kappa}_{i_\kappa}\rangle$
  and $I\equiv(i_1,...,i_S)$, while the last two terms refer to the expansion over the species and particle
  SPFs. We introduce the single-hole function (SHF) $|\tilde{\Psi}^{\alpha;\kappa}_i\rangle$ for each SPF as
  $|\Psi\rangle=:\sum_i|\phi^{\alpha;\kappa}_i\rangle|\tilde{\Psi}^{\alpha;\kappa}_i\rangle$ with $i$ summing
  over all the SPFs of the related node, and the SHFs of the species and particle layer are given as 
  \begin{align} \label{SHF1}
  |\tilde{\Psi}^{2;\kappa}_j\rangle &= \sum_{I^\kappa}A^1_{I^\kappa_j}\prod_{\mu\neq\kappa}|\phi^{2;\mu}_{i_\mu}\rangle,\\\label{SHF2}
  |\tilde{\Psi}^{3;\kappa}_j\rangle_{\overline{l}} &= \sum_i\sum_{\vec{n}_\kappa|N_\kappa-1}\sqrt{\frac{n_j+1}{N_\kappa}}A^{2;\kappa}_{i;\vec{n}_\kappa+\hat{j}}|\vec{n}_\kappa\rangle_{\overline{l}}|\tilde{\Psi}^{2;\kappa}_i\rangle.
  \end{align}
  In equation (\ref{SHF1}), $I^\kappa$ is defined as an integer array of $(i_1,i_2,...,i_{\kappa-1},i_{\kappa+1},...,i_s)$,
  $i.e.$, taking the index of $i_\kappa$ out of $I$, and $I^\kappa_j$ is defined as
  $(i_1,i_2,...,i_{\kappa-1},j,i_{\kappa+1},...,i_s)$. Comparing equations (\ref{delta},\ref{partial}),
  the time derivative of $|\Psi\rangle$ has the same overall structure as $|\delta\Psi\rangle$.
  
  The equations of motion are derived by substituting equations (\ref{delta},\ref{partial}) into the
  Dirac-Frenkel variational principle and applying the restriction of
  $\langle\phi^{\alpha;\kappa}_{i}|\partial_t|\phi^{\alpha;\kappa}_j\rangle=0$, for ensuring the
  orthonormality of the SPFs. In equations (\ref{delta},\ref{partial}) we may only focus exclusively
  on $l=1$ in the summation over $l$ in the last term, as for each species $\kappa$ the terms of
  different $l$ are identical to each other due to the indistinguishability of the bosons belonging
  to the same species. Due to the same reason, we drop the subscripts of $l$ and $\overline{l}$ for the
  permanents and SHFs in the following. The equations of motion for the coefficients read as 
  \begin{align}\label{eqn_top}
  i\partial_t A^1_I &= \sum_J \langle I|H|J\rangle A^1_J,\\ \label{eqn_nontop}
  i\partial_t A^{\alpha;\kappa}_{n;C}&=\sum_{m,D}\sum_{p} \langle C^{\alpha;\kappa}|(1-\hat{P}^{\alpha;\kappa})(\rho^{\alpha;\kappa})^{-1}_{n,p}\langle\hat{H}\rangle^{\alpha;\kappa}_{p,m}|D^{\alpha;\kappa}\rangle A^{\alpha;\kappa}_{m;D}.
  \end{align}
  In equation (\ref{eqn_nontop}) $|C^{\alpha;\kappa}\rangle$ and $|D^{\alpha;\kappa}\rangle$ belong to the
  basic basis of the corresponding node, $i.e.$, the basis used to build up the
  SPFs of the related node. For the species nodes, the basic basis is the permanent state basis
  $\{|\vec n_\kappa\rangle|N_\kappa\}$, and that of a particle node is $\{|r^\kappa_j\rangle\}_{j=1}^{\mathscr{M}_\kappa}$.
  The projection operator $\hat{P}^{\alpha;\kappa}$ is then defined as
  $\hat{P}^{\alpha;\kappa}=\sum_{i,E,F}(A^{\alpha;\kappa}_{i;F})^*A^{\alpha;\kappa}_{i;E}|E^{\alpha;\kappa}\rangle\langle F^{\alpha;\kappa}|$.
  
  We also introduce in equation (\ref{eqn_nontop}) the concepts of density matrices and mean-field operators, as follows 
  \begin{align}\label{dmat0}
  (\rho^{\alpha;\kappa})_{i,j}&=\langle\tilde{\Psi}^{\alpha;\kappa}_i|\tilde{\Psi}^{\alpha;\kappa}_j\rangle;\\\label{mf2}
   \langle\hat{H}\rangle^{2;\kappa}_{i,j}&=\sum_{\vec{n}_\kappa,\vec{m}_\kappa}\langle\tilde{\Psi}^{2;\kappa}_{i}|\langle\vec{n}_\kappa|\hat{H}|\vec{m}_\kappa\rangle|\tilde{\Psi}^{2;\kappa}_{j}\rangle|\vec{n}_\kappa\rangle\langle\vec{m}_\kappa|;\\\label{mf3}
   \langle\hat{H}\rangle^{3;\kappa}_{i,j}&=\sum_{r^\kappa_1,r^\kappa_2}\langle\tilde{\Psi}^{3;\kappa}_{i}|\langle r^\kappa_1|\hat{H}|r^\kappa_2\rangle
   |\tilde{\Psi}^{3;\kappa}_{j}\rangle|r^\kappa_1\rangle\langle r^\kappa_2|.
  \end{align}
  Please note that $(\rho^{\alpha;\kappa})^{-1}_{n,p}$ in equation (\ref{eqn_nontop}) refers to the (n,p)-th element of the inverse of the regularized density matrix (\ref{dmat0}) \cite{BJWM00}.
  The above equations demonstrate the equations of motion for the coefficients in all layers of species ML-MCTDHB, and with these equations the time evolution of the system can be deduced.
  
  Equations (\ref{eqn_top},\ref{eqn_nontop}) give the general form of the equations of motion for species ML-MCTDHB,
  as well as the particle ML-MCTDHB, which will be introduced in section \ref{subsec_par}. It turns out that the equations of motion
  for ML-MCTDHB take the same form of ML-MCTDH, despite that the two methods are dealing with indistinguishable and distinguishable
  particles, respectively. The difference between the two methods comes in the calculation of the ingredients of the equations, $i.e.$,
  the density matrices and mean-field operators introduced in equations (\ref{dmat0}-\ref{mf3}). The formal similarity of the equations of motion
  of ML-MCTDHB and ML-MCTDH offers the possibility to merge the two methods together, and deal with mixtures consisting of distinguishable and indistinguishable particles.
 
\subsubsection{Details and Construction of the Ingredients for the Equations of Motion}

  In the preceding section we have introduced the general form of the equations of motion with all necessary ingredients in equations (\ref{eqn_top}-\ref{mf3}).
  In this section we present the detailed expressions for the equations on different layers, with a focus on the second-quantization treatment
  on the bosonic species layer and particle layer. As this section mainly supplies the explicated equations of motion, it will not affect 
  the understanding of the whole method to skip the detailed expressions in this section. Here we consider a general Hamiltonian (\ref{Ham_1st}) containing single-particle terms,
  as well as two-body interaction terms. We adapt the product form representation of the interaction potentials, rewriting the intra- and inter-species interactions as 
  \begin{align}\label{potfit}
   V_\kappa(r^\kappa_1,r^\kappa_2) &= \sum^{\mathscr{V}_{\kappa}}_{n=1} C^\kappa_n V^\kappa_{n,1}(r^\kappa_1)V^\kappa_{n,2}(r^\kappa_2),\\\nonumber
   W_{\kappa_1\kappa_2}(r^{\kappa_1}_1,r^{\kappa_2}_2) &= \sum^{\mathscr{W}_{\kappa_1,\kappa_2}}_{n=1} D^{\kappa_1,\kappa_2}_n W^{\kappa_1\kappa_2}_{n,\kappa_1}(r^{\kappa_1}_1)W^{\kappa_1\kappa_2}_{n,\kappa_2}(r^{\kappa_2}_2).
  \end{align}
  In (\ref{potfit}), $V^\kappa_{n,p}$ and $W^{\kappa_1\kappa_2}_{n,\kappa_p}$ ($p=1,2$) form the expansion basis of
  the product representations of the intra- and inter-species interactions $V_\kappa(r^\kappa_1,r^\kappa_2)$ and
  $W_{\kappa_1\kappa_2}(r^{\kappa_1}_1,r^{\kappa_2}_2)$, respectively. $C^\kappa_n$ and $D^{\kappa_1,\kappa_2}_n$
  are the expansion coefficients for the intra- and inter-species interactions. Please note that one can bring any
  interaction potential to the product form (\ref{potfit}). A convenient way of choosing the basis for the one-body
  operators $V^\kappa_{n,p}$ and $W^{\kappa_1\kappa_2}_{n,\kappa_p}$ ($p=1,2$) is given by the algorithm POTFIT \cite{POTFIT1,POTFIT2},
  by using the basis of the so-called natural potentials. The natural potential, for instance, $V^\kappa_{n,1}$ is defined as the
  eigenvectors of the potential density matrix
  $(V^\kappa)_{ij}=\sum_{r^\kappa=1}^{\mathscr{M}_\kappa}V_\kappa(r^\kappa_i,r^\kappa)^*V_\kappa(r^\kappa_j,r^\kappa)$.
  Consequently, $C^\kappa_n$ is the overlap between $V_\kappa$ and the product $V^\kappa_{n,1}V^\kappa_{n,2}$.
  The product form of the interactions allows the equations of motion to be expressed in a neat manner, and can
  improve the numerical performance if only a few terms are sufficient for a fair representation of the interaction potentials.
  
  In ML-MCTDHB, the SPFs basis of each layer are truncated with respect to the lower layer, and the operators composing
  the Hamiltonian (\ref{Ham_1st}) can be expanded in the SPFs basis of each layer. Now we demonstrate the operator expansion
  with respect to the SPFs basis of each layer, and for the particle layer it is convenient to use the second quantization
  representation for the operators. The single particle operators, with respect to the equation of motion (\ref{eqn_nontop})
  for $\alpha=3$, including the single particle Hamiltonian terms and $\hat{W}^{3;\kappa_1,\kappa_2}_{n,\kappa_p}$ $(p=1,2)$
  for inter-species interaction terms, are given on the particle layer as follows
  \begin{align}\label{ham_single}
   \hat{H}^{3;\kappa}_0&=\sum_{i,j}(H^{3;\kappa}_0)_{ij}\hat{a}^\dagger_{\kappa,i}\hat{a}_{\kappa,j} \\\nonumber
   \hat{W}^{3;\kappa_1,\kappa_2}_{n,\kappa_p}&=\sum_{i,j}(W^{3;\kappa_1,\kappa_2}_{n,\kappa_p})_{ij}\hat{a}^\dagger _{\kappa_p,i}\hat{a}_{\kappa_p,j}.  \quad\quad (p=1,2)
  \end{align}
  Here $\hat{a}_{\kappa,i}$ ($\hat{a}^\dagger _{\kappa,i}$) refers to the operator annihilating (creating)
  a $\kappa$ boson in the SPF state $|\phi^{3;\kappa}_i\rangle$. $(H^{3;\kappa}_0)_{ij}$ is defined as
  $(H^{3;\kappa}_0)_{ij}\equiv\langle \phi^{3;\kappa}_i|\hat H^0_\kappa|\phi^{3;\kappa}_j\rangle$, and
  $(W^{3;\kappa_1,\kappa_2}_{n,\kappa_p})_{ij}$ follows the same definition. The two-particle operators
  for the intra-species interactions are given as 
  \begin{align}\label{ham_intra}
   \hat{V}^{3;\kappa} &= \frac{1}{2}\sum_{i_1,i_2,j_1,j_2}(V^{3;\kappa})_{i_1i_2j_1j_2}\hat{a}^\dagger _{\kappa,i_1}\hat{a}^\dagger _{\kappa,i_2}\hat{a}_{\kappa,j_1}\hat{a}_{\kappa,j_2}\\\nonumber
   &=\frac{1}{2}\sum_{n}C^\kappa_n\sum_{i_1,i_2,j_1,j_2}(V^{3;\kappa}_{n,1})_{i_1j_1}(V^{3;\kappa}_{n,2})_{i_2j_2}\hat{a}^\dagger _{\kappa,i_1}\hat{a}^\dagger _{\kappa,i_2}\hat{a}_{\kappa,j_1}\hat{a}_{\kappa,j_2},
  \end{align}
  with $(V^{3;\kappa}_{n,p})_{ij}=\langle \phi^{3;\kappa}_i|\hat V^{\kappa}_{n,p}|\phi^{3;\kappa}_j\rangle$ ($p=1,2$).
  
  On the species layer, the terms of the single particle Hamiltonian become 
  \begin{align}\label{ham_h0}
    \hat{H}^{2;\kappa}_0&=\sum_{ij}(H^{2;\kappa}_0)_{ij}|\phi^{2;\kappa}_i\rangle\langle\phi^{2;\kappa}_j|,\\\nonumber
    (H^{2;\kappa}_0)_{ij}&\equiv\sum_{p,q}(H^{3;\kappa}_0)_{pq}(\tilde{Q}^\kappa_{pq})_{ij}.
  \end{align}
  The intra-species interaction operators also become effectively single particle operators with respect to the species SPFs, as 
  \begin{align}\label{ham_v}
  \hat{V}^{2;\kappa}&=\sum_{ij}(V^{2;\kappa})_{ij}|\phi^{2;\kappa}_i\rangle\langle\phi^{2;\kappa}_j|,\\\nonumber
  (V^{2;\kappa})_{ij}&\equiv\frac{1}{2}\sum_{p_1p_2q_1q_2}(V^{3;\kappa})_{p_1p_2q_1q_2}(\tilde{P}^\kappa_{p_1p_2q_1q_2})_{ij}.
  \end{align}
  In equations (\ref{ham_h0},\ref{ham_v}), we define $(\tilde Q^{\kappa}_{pq})_{ij}=\sum_{\vec{n}|N_\kappa-1}(A^{2;\kappa}_{i;\vec{n}+\hat{p}})^*A^{2;\kappa}_{j;\vec{n}+\hat{q}}Q_{\vec{n}}(p,q)$ and $(\tilde P^{\kappa}_{p_1p_2q_1q_2})_{ij} = \sum_{\vec{n}|N_\kappa-2}(A^{2;\kappa}_{i;\vec{n}+\hat{p}_1+\hat{p}_2})^*A^{2;\kappa}_{j;\vec{n}+\hat{q}_1+\hat{q}_2}P_{\vec{n}}(p_1,p_2)P_{\vec{n}}(q_1,q_2)$. The summation index $\vec{n}|N_\kappa-k$ refers to all the permanent states of $(N-k)$ bosons, while $\vec{n}+\hat{q}$ and $\vec{n}+\hat{q}_1+\hat{q}_2$ refers to adding one boson to the $q$-th SPF state of $\vec n$ and adding two bosons to the $q_1$-th and $q_2$-th SPF states of $\vec{n}$, respectively. We also introduce here $P_{\vec{n}}(l,m)=\sqrt{(n_l+1+\delta_{l,m})(n_m+1)}$ and $Q_{\vec{n}}(i,j)=\sqrt{(n_i+1)(n_j+1)}$. The single particle operators $\hat{W}^{2;\kappa_1,\kappa_2}_{n,p}$, are defined in the same way as the single particle Hamiltonian terms in 
equation (\ref{ham_h0}).
  
  Then we can show the detailed form of the equations of motion for the coefficients on all layers. The top layer equations of motion turn out to be 
  \begin{align}\label{eqom_top2}
  i\partial_t A^1_I &= \sum_\kappa\sum_j [(H^{2;\kappa}_0)_{i_\kappa j}+(V^{2;\kappa})_{i_\kappa j}] A^1_{I^\kappa_j}\\ \nonumber
  &+\sum_{\kappa_1<\kappa_2}\sum_n\sum_{j_1,j_2}D^{\kappa_1,\kappa_2}_n(W^{2;\kappa_1,\kappa_2}_{n,\kappa_1})_{i_{\kappa_1}j_1}(W^{2;\kappa_1,\kappa_2}_{n,\kappa_2})_{i_{\kappa_2}j_2}A^1_{I^{\kappa_1\kappa_2}_{j_1j_2}}.   
  \end{align}
  $I$ is again defined as an array of $(i_1,i_2,...,i_S)$, while $I^\kappa_j$ and $I^{\kappa_1\kappa_2}_{j_1j_2}$ are the arrays obtained by replacing
  $i_\kappa$ in $I$ by $j$ and $i_{\kappa_1}$, $i_{\kappa_2}$ in $I$ by $j_1$, $j_2$, respectively.
  
  The equations of motion for the coefficients on the non-top layers can be obtained by substituting the related density matrices
  and mean-field operators to equation (\ref{eqn_nontop}). The density matrices on the species and particle layers are calculated, respectively, as 
  \begin{align}\label{dmat}
    (\rho^{2;\kappa})_{ij}&=\sum_{I^\kappa}(A^1_{I^\kappa_i})^*A^1_{I^\kappa_j},\\\nonumber
    (\rho^{3;\kappa})_{ij}&=\frac{1}{N_\kappa}\sum_{kl}\rho^{2;\kappa}_{kl}(\tilde Q^\kappa_{ij})_{kl}.
  \end{align}
  On the species layer the mean-field operators for the terms of the single particle Hamiltonian and intra-species
  interactions are calculated, in the second quantization picture, as 
  \begin{align}\label{mf_spc_intra}
   \langle \hat{H}^0_\kappa+\hat{V}_\kappa\rangle^{2;\kappa}_{ij}&=(\rho^{2;\kappa})_{ij}(\hat{H}^{3;\kappa}_0+\hat{V}^{3;\kappa}).
  \end{align}
  The mean-field operator for all inter-species interactions acting on the $\kappa$ species is given by
  $\langle \hat W_\kappa\rangle^{2;\kappa}_{ij}=\sum_{\nu\neq\kappa}\langle \hat W_{\kappa}\rangle^{2;\kappa,\nu}_{ij}$,
  where the mean field operator for the inter-species interaction between the $\kappa$ and $\nu$ species is obtained by
  $\langle \hat W_{\kappa}\rangle^{2;\kappa,\nu}_{ij}=\sum_{n=1}^{\mathscr{W}_{\kappa,\nu}}D^{\kappa,\nu}_n\langle \hat{W}_{n,\kappa}\rangle^{2;\kappa,\nu}_{ij}$ with
  \begin{align}\label{mf_spc_inter}
   \langle \hat{W}_{n,\kappa}\rangle^{2;\kappa,\nu}_{ij}&=\hat{W}^{3;\kappa,\nu}_{n,\kappa}\cdot (\tilde{W}^{2;\kappa,\nu}_{n,\kappa})_{ij},\\\nonumber
   (\tilde{W}^{2;\kappa,\nu}_{n,\kappa})_{ij}&=\sum_{I^{\kappa,\nu}}\sum_{rs}(A^1_{I^{\kappa,\nu}_{i,r}})^*A^1_{I^{\kappa,\nu}_{j,s}}(W^{2;\kappa,\nu}_{n,\nu})_{rs}.
  \end{align}
  
  On the particle layer, the mean-field operators for the terms of the single particle Hamiltonian, the intra-species interactions and the inter-species interactions are calculated by 
  \begin{align}\label{mf_par}
   \langle \hat H^0_\kappa\rangle^{3;\kappa}_{ij}&=(\rho^{3;\kappa})_{ij}\hat{H}^0_\kappa,\\\nonumber
   \langle \hat V_\kappa\rangle^{3;\kappa}_{ij}&=\frac{1}{N_\kappa}\sum_{n=1}^{\mathscr{V}_{\kappa}}C^\kappa_n \langle \hat V_{n;\kappa}\rangle^{3;\kappa}_{ij},\\\nonumber
   \langle \hat W_\kappa\rangle^{3;\kappa}_{ij}&=\frac{1}{N_\kappa}\sum_{\nu\neq\kappa}\sum_{n=1}^{\mathscr{W}_{\kappa,\nu}}D^{\kappa,\nu}_n\langle \hat W_{n,\kappa}\rangle^{3;\kappa,\nu}_{ij},
   \end{align}
   with the mean field operators $\langle \hat V_{n;\kappa}\rangle^{3;\kappa}_{ij}\equiv(\hat{V}^\kappa_{n,1})\big(\sum_{rs}(\rho^{2;\kappa})_{rs}\sum_{pq}(\tilde P^\kappa_{ipjq})_{rs}(V^{3;\kappa}_{n,2})_{pq}\big)$,
   and $\langle \hat W_{n,\kappa}\rangle^{3;\kappa,\nu}_{ij}=(\hat W^{\kappa,\nu}_{n,\kappa})(\sum_{rs}(\tilde W^{2;\kappa,\nu}_{n,\kappa})_{rs}(\tilde Q^{\kappa}_{ij})_{rs})$.
   Please note the recursive character of the formulas for the mean-field operators and the reduced density matrices,
   which is a consequence of the product representation of the interaction potentials. The equations of motion without
   the product representation of the interactions are given in the appendix \ref{spc_mctdhb}.

\subsection{ML-MCTDHB for High-dimensional Bosonic Systems}\label{subsec_par}
\begin{figure}
 \centering
 \subfigure{\thesubfigure
 \label{fig_part_ml_blackbox}
 \includegraphics[width=0.3\textwidth]{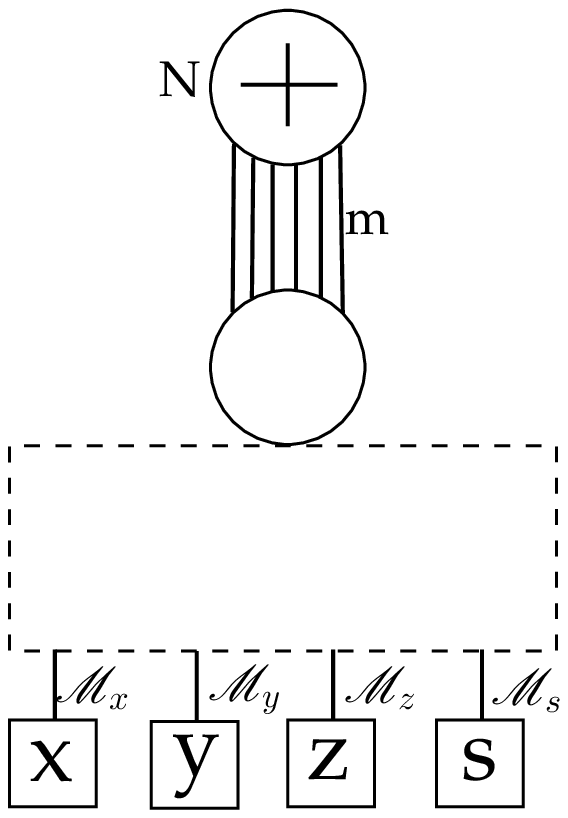}
}
  \hspace{1in}
\subfigure{\thesubfigure
\label{pmlb_tree}
\includegraphics[width=0.3\textwidth]{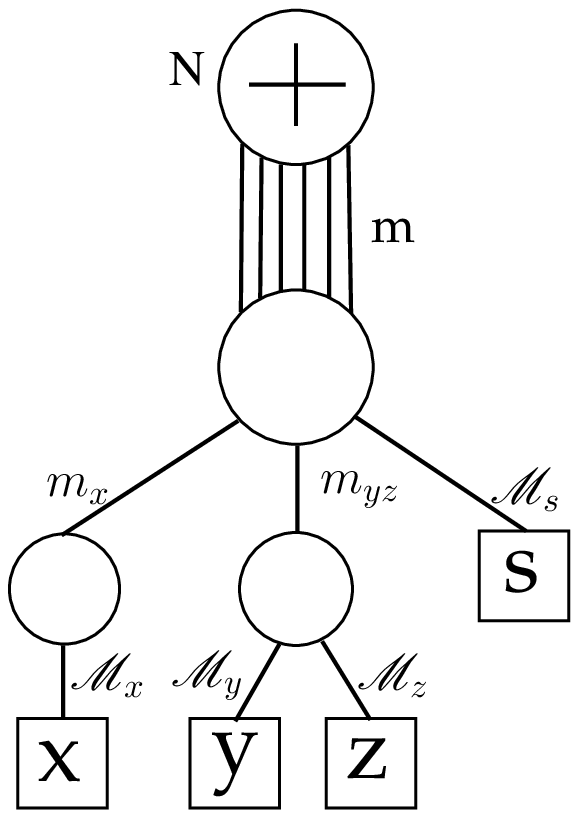}

}
\caption{ (a) An illustration of a general particle multi-layer MCTDHB wave function
expansion for spin-one bosons in three-dimensional coordinate space, with $\mathscr{M}_s=3$ in the figures. The dashed
box indicates the various possibilities for mode combination of the primitive
DOF, i.e. the spatial DOF x, y, z and the spin projection on the z-axis.
(b) An example of the mode combination scheme inside the dashes box of (a),
where the spatial DOF y and z are combined into one logical DOF. }

\end{figure}

  Now we proceed to demonstrate the multi-layer treatment of high-dimensional bosonic systems, which is termed
  as \textit{particle multi-layer MCTDHB} (particle ML-MCTDHB). Particle ML-MCTDHB has its origin in the mode
  combination of MCTDH \cite{HDMey09}, as well as the multi-layer scheme of ML-MCTDH \cite{Manth08}. In this
  section we take a single-species bosonic system in high-dimensional space as an example to demonstrate the
  particle ML-MCTDHB, and the combination of species and particle ML-MCTDHB will be given in the next section.
  This example also shows how ML-MCTDHB deals with a single-species system, which is also a supplement to the
  general discussion of bosonic mixtures in the previous sections. We consider a single-species bosonic system
  of N identical bosons living in a Q-dimensional configuration space, spanned by the primitive DOF
$\{x_1,x_2,...,x_Q\}$. In ML-MCTDHB, we firstly perform a mode combination of
these primitive DOF, and generate the logical DOF one layer above, as 
\begin{align}\label{md-com}
 x^{P-1}_{1}&=x^{P-1}_{1}(x_{1},x_{2},...,x_{n_{1}}),\\\nonumber
 x^{P-1}_{2}&=x^{P-1}_{2}(x_{n_{1}+1},x_{n_{1}+2},...,x_{n_{1}+n_{2}}),
		    \\\nonumber
	    &\;\;\vdots \\\nonumber
 x^{P-1}_{N_{P-1}}&=x^{P-1}_{N_{P-1}}(x_{N-n_{N_{P}}+1},x_{N-n_{N_{P}}+2},...,x_Q).
\end{align}
Such mode combination can be repeated
recursively layer by layer, until all the DOF are combined into a single logical DOF, which correspond to the particle node.
Above this particle node, there is the top node representing the state vector $|\Psi\rangle$ of the
whole system. This mode-combination procedure corresponds to the tree structure
shown in figure \ref{fig_part_ml_blackbox}, where the bottom nodes are the so-called primitive nodes, containing the 
primitive DOF, and the dashed box between the
particle node and the primitive nodes indicates the variety of possible
mode-combination schemes.

For each primitive DOF, a set of time-independent basis vectors are assigned.
Above the primitive nodes, a set of time-dependent SPFs is associated to each
logical DOF on the non-top layer, which are expanded with respect to the SPFs
of the corresponding DOF one layer below, as 
\begin{equation}\label{MLB_spf_exp}
|\phi^{\alpha;\kappa}_{n}\rangle=\sum_{i_{1},i_{2},...,i_{m}}A^{\alpha;\kappa
}_{
n;i_{1},i_{2},...,i_{m}} \prod_{r\in M^{\alpha;\kappa}}
|\phi^{\alpha+1;\kappa_{r}}_{i_{r}}\rangle
\equiv \sum_{I}A^{\alpha;\kappa}_{n;I}|I^{\alpha;\kappa}\rangle.
\end{equation}
Here $|\phi^{\alpha;\kappa}_{n}\rangle$ is the $n$-th SPF of the
$\kappa$-th DOF on the $\alpha$-th layer, and this logical DOF is a combination
of a set $M^{\alpha;\kappa}$ of m DOF of the lower layer, where $M^{\alpha;\kappa}$
keeps track of the correct subnode indices $\kappa_r$ belonging to the node
$(\alpha;\kappa)$. The state vector $|\Psi\rangle$ of the
top node is expanded with respect to permanents 
\begin{equation}
|\Psi\rangle=\sum_{\vec{n}|N}A^1_{\vec{n}}|\vec{n}\rangle.
\end{equation}
 An example for incorporating primitive mode combination in particle ML-MCTDHB
 is shown in figure \ref{pmlb_tree}.The tree structure in figure \ref{pmlb_tree}
 corresponds to a single-species bosonic ensemble, and the bosons possess four primitive DOF,
 $i.e.$, the three spatial DOF along x, y and z directions as well as an internal DOF,
 $e.g.$, a spin DOF. In the mode combination shown here, the y and z DOF
 are grouped into one logical DOF, and the primitive basis of x DOF is truncated
 to a set of SPF, associated with a logical DOF. Then these two logical DOF are combined
 with the internal primitive DOF to the upper layer logical DOF represented by the
 particle node. Physically this mode combination might correspond, as an example,
to the strongly correlated $y$ and $z$ DOF with weak correlation to the $x$ DOF.

For simplicity, we restrict ourselves to two-body interactions and, moreover,
assume the product form representations of the Hamiltonian as usual
\begin{align}\label{part_ml_hamil}
 \hat{H}&=\sum_{i}\hat{h}(i)+\sum_{i<j}\hat{V}(i,j) \\\nonumber
&=\sum_i\hat{h}(i)+\sum_{i<j}\sum_\nu C_\nu\hat{v}^1_\nu(i)\hat{v}^2_\nu(j).
\end{align}
$\hat{h}(i)$ and $\hat{V}(i,j)$ are the one-body Hamiltonian acting on the
$i$-th boson, and the two-particle interaction operator acting on the $i$-th and
$j$-th bosons, respectively, where the two-particle interaction operator is
written as the summation of products of single particle operators acting on the
$i$-th and $j$-th boson separately.

The equations of motion of all the coefficients are again obtained by the
Dirac-Frenkel variational principle and the equations for the top and particle nodes read 
\begin{align} \label{eq_parml}
 i\partial_t A^1_{\vec{n}}&=\sum_{\vec{m}}\langle\vec{n}|\hat{H}|\vec{m}\rangle
A^1_{\vec{m}}, \\\nonumber
 i\partial_t A^{2;1}_{n;I}&=\sum_J\langle
I^{2;1}|(1-\hat{P}^{2;1})\hat{h}|J^{2;1}\rangle A^{2;1}_{n;J}\\\nonumber 
&\qquad+\sum_{\nu}C_\nu
\{\sum_{J}\langle I^{2;1}|(1-\hat{P}^{2;1})\hat{v}_{\nu;1}|J^{2;1}\rangle\}\\\nonumber
&\qquad\times\{\sum_{m}[(\rho^{2;1})^{-1}\times\langle
\tilde{v}^{2;1}_{\nu;2}\rangle]_{n,m}\} A^{2;1}_{m;J}. 
\end{align}
For the particle node,
$\hat{P}^{2;1}\equiv\sum_{n}|\phi^{2;1}
_n\rangle\langle\phi^{2;1}_n|$ is the projection operator, and for the single species
case there is only one node on the particle layer, we have $\kappa=1$ on this layer. The 
density matrix $\rho^{2;1}$ is calculated as 
\begin{equation}\label{dmat_pml}
 \rho^{2;1}_{ij}=\frac{1}{N}\sum_{\vec{n}|N-1}Q_{\vec n}(i,j)(A^1_{\vec{n}
+\hat{i}})^*A^1_{\vec{n}+\hat{j}},
\end{equation}
and the mean-field matrices $\langle\tilde{v}^{2;1}_{\nu;2}\rangle$ are
calculated
as 
\begin{align}\label{mf_pot_inter}
 \langle
\tilde{v}^{2;1}_{\nu;2}\rangle_{i,j}&=\frac{1}{N}\sum_{\vec{n}|N-2}\sum_{p,q}P_{\vec n}(i,p)P_{\vec n}(j,q) (A^1_{\vec{n}+\hat{i}+\hat{p}})^*A^1_{\vec{n}+\hat{j}
+\hat{q }}
\langle\phi^{2;1}_p|\hat{v}^2_{\nu}|\phi^{2;1}_q\rangle.
\end{align}

Assuming that the one-body operators $\hat{h}$, $\hat{v}^i_\nu$ (i=1,2) are also given in a
product form with respect to the primitive DOF, these operators can be expressed with respect
to the logical DOF of each layer on the layers below the particle layer (cf. \cite{Manth08}), e.g.
\begin{equation}
 \hat{h}^{\alpha;\kappa}=\sum_{\mu}D_\mu\prod_{\kappa}\hat{h}^{\alpha;\kappa}_\mu,
\end{equation}
Here $\hat{h}^{\alpha;\kappa}_\mu$ is an operator acting on the $(\alpha;\kappa)$ DOF, and
can be obtained recursively by 
$(\hat{h}^{\alpha;\kappa}_\mu)_{ij}=\langle\phi^{\alpha;\kappa}_i|\prod_{\nu\in M^{\alpha;\kappa}}\hat{h}^{\alpha+1;\nu}_\mu|\phi^{\alpha;\kappa}_j\rangle$.
Then the density and mean-field matrices of the logical DOF below the particle layer are
generated the same way as in ML-MCTDH, which enables one to apply the well-known recursion
schemes \cite{Manth08,VHDM11}. The equations of motion can then be obtained by substituting
the related operators, density matrices and mean-field matrices to equation (\ref{eqn_nontop}),
which turns out to be general for both the species and particle ML-MCTDHB.

\subsection{ML-MCTDHB for General Bosonic Systems}\label{subsec_intg}
\begin{figure}
\includegraphics[width=0.6\textwidth]{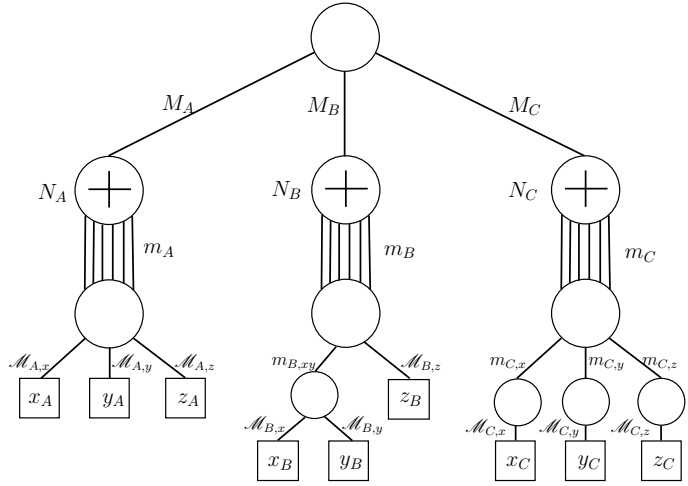}
\caption{A tree diagram of the combined treatment of the species and particle ML-MCTDHB for a
three-species bosonic system. The three species are again labeled as A, B and C bosons as in Figure
\ref{tree_sketch}. In contrast to that example, the bosons of all species may move in three-dimensional space here.  }
\label{tree_sketch2}
\end{figure}

In this section we demonstrate the general ML-MCTDHB theory which is a combination of the
species ML-MCTDHB and particle ML-MCTDHB. The general ML-MCTDHB approach extends our study
to bosonic mixtures in high-dimensional systems and even mixed-dimensional systems. As an
example, we consider a three-species bosonic mixture in three-dimensional space, and the three
species are again named as A, B and C bosons. Firstly a tree diagram related to this
three-species mixture is given in figure \ref{tree_sketch2}. This three-species mixture is
different from the mixture in figure \ref{tree_sketch} by the fact that the A and B bosons
have three primitive DOF in figure \ref{tree_sketch2}. Despite the difference,
the tree diagrams for both mixtures have the same structure from top to particle layer, which
indicates that the setup on the particle multi-layer level will not affect the species multi-layer
level, which is above the particle layer. On the particle multi-layer level, the three 
species follow different mode combination schemes, determined by the correlations between the DOF
of each boson.
The different mode combinations of the three species in figure \ref{tree_sketch2} indicate that the
three DOF of A bosons are all strongly correlated, while for B bosons only the x and y DOF are strongly
correlated, which are combined together, and the three DOF of C bosons have loose correlations between
each other. The ansatz for A, B and C bosons on the particle multi-layer level can be obtained following
the discussion in section \ref{subsec_par}. To conclude, the species and particle ML-MCTDHB are actually
independent of each other, and they can be straightforwardly connected on the particle layer. The structure
of the species level only depends on the correlations between the bosonic species, while the structure on
the particle level is determined by the correlations of the primitive DOF.
 
Similar to the ansatz for the combination of species and particle ML-MCTDHB, the equations
of motion are compatible with each other. A brief sketch of the link between species and
particle ML-MCTDHB goes as follows: At each time step the Hamiltonian operators are generated
for each layer from primitive to top layer, then the density matrices and mean-field operators
are constructed from top to bottom. The calculation of the Hamiltonian operators, density matrices
and the mean-field operators on a certain layer follows the corresponding rules given in
sections \ref{subsec_species} and \ref{subsec_par}.
Substituting all the components into the corresponding equations, the right hand side (rhs) of the
equations of motion is ready to be handed to the integrator.

For the calculation of the equations, one point not mentioned in section \ref{subsec_par} are the
mean-field operators of the interspecies interactions on the particle multi-layer level. For this point,
we consider a general case of interspecies interaction between the $\kappa_1$ and $\kappa_2$ bosons.
On the $\alpha$-th layer below the particle layer this interaction is written in the POTFIT form as
$\hat W^{\alpha;\kappa_1,\kappa_2}=\sum_n D^{\alpha;\kappa_1,\kappa_2}_{n} \big(\prod_{i=1}^{m^{\alpha,\kappa_1}} \hat W^{\alpha;\kappa_1,\kappa_2}_{n,\kappa_1,i}\big)\big(\prod_{j=1}^{m^{\alpha,\kappa_2}} \hat W^{\alpha;\kappa_1,\kappa_2}_{n,\kappa_2,j}\big)$,
where there are $m^{\alpha,\kappa_1}$ and $m^{\alpha,\kappa_2}$ DOF for the $\kappa_1$ and $\kappa_2$
bosons on the $\alpha$ layer. Assuming that the mean-field operator for the interspecies interaction on
the particle layer has been calculated by equation (\ref{mf_pot_inter}), the mean-field operator on the
$\alpha$-layer below the particle layer can be calculated as 
\begin{align}
 \langle \hat W^{\alpha;\kappa_1,\kappa_2}_{n,\kappa_1,a}\rangle_{ij}&=(\hat W^{\alpha+1;\kappa_1,\kappa_2}_{n,\kappa_1,a})\langle \tilde W^{\alpha;\kappa_1,\kappa_2}_{n,\kappa_1,a}\rangle_{ij},\\\nonumber
 \langle \tilde W^{\alpha;\kappa_1,\kappa_2}_{n,\kappa_1,a}\rangle_{ij}&=\sum_{rs}\langle \tilde W^{\alpha-1;\kappa_1,\kappa_2}_{n,\kappa_1,b}\rangle_{rs}\sum_{I^a}\sum_{J^a}(A^{\alpha-1;\kappa_1}_{r;I^a_i})^*A^{\alpha-1;\kappa_1}_{s;J^a_j}\prod_{p\in M^{\alpha-1;\kappa_1,b} \atop p\neq a}(W^{\alpha,\kappa_1,\kappa_2}_{n,\kappa_1,p})_{i_p,j_p},
\end{align}
where $M^{\alpha-1;\kappa_1,b}$ is the set of $\alpha$-layer DOF combined into the $b$-th DOF of the
$\kappa_1$ species and $(W^{\alpha,\kappa_1,\kappa_2}_{n,\kappa_1,p})_{i_p,j_p}$ are the matrix elements of the corresponding operator.
For a more detailed description of the mean-field operators we refer to \cite{Manth08}.

The general ML-MCTDHB method can deal with bosonic mixtures in high-dimensional space and even mixed
dimensions, and a major advantage of the method lies in its flexibility with respect to the SPFs of each DOF.
The method can handle both the weak and strong correlation regimes, simply by adjusting the number of the SPFs
of related DOF. For instance, when the intra- and inter-species interactions are weak, we can take few particle and species
SPFs, and the extreme case is that we take only one particle and one species SPF for each species, which reduces to
the mean-field treatment. On the other hand, in the case of strong intra-species interaction but weak inter-species
interaction, we need to supply more particle SPFs, but few species SPFs. One can also reach the full CI limit by supplying
as many SPFs for each DOF as the basic basis states of the corresponding DOF.

\subsection{Scaling}\label{subsec_scaling}
At this stage, we would like to point out the relationship between ML-MCTDHB
and related methods. For a single-species system,
ML-MCTDHB and MCTDHB \cite{ASC07,ASC08,SAC10} coincide for one-dimensional bosons without internal
degrees of freedom. When considering a single species of bosons living in a
higher-dimensional configuration space, $e.g.$ with both spatial and internal DOF, ML-MCTDHB reduces to MCTDHB if all the
primitive modes are directly combined to the particle node. 
For bosonic mixtures, MCTDHB has been generalized to deal with mixtures, $e.g.$,
containing two or three bosonic species, with methods known as MCTDH-BB \cite{MCTDHBB}
and MCTDH-BBB \cite{ASSC11}, respectively.
In the following we categorize them as standard MCTDHBs in comparison to the Multi-Layer
MCTDHB.
The state vector of the total system is expanded in standard MCTDHBs as the summation
over the products of the permanents states belonging to each species, while in
species multi-layer MCTDHB the state vector is firstly expanded with respect to
Hartree products of species SPFs, and the species SPFs are then expanded within
the permanents basis of each species. Hence, ML-MCTDHB recovers the
standard MCTDHBs treatment if as many species SPFs are provided
as there are configurations for each species.

We finally compare the scaling of the methods MCTDH, ML-MCTDH, standard MCTDHBs and
ML-MCTDHB applied to this multi-species setup by investigating the memory
consumption for storing the total wave function. This also provides us with a
rough estimate for the performances of the methods. Let us consider a bosonic mixture
of $S$ species, of which each species contains $N$ bosons, and  we denote the number of
grid points and particle SPFs by $n$ and $m$, respectively. If we neglect the symmetrization option as well as the
possibility of primitive mode combination, $m^{SN}+SNmn$ coefficients
have to be propagated in MCTDH. The ML-MCTDH expansion equivalent to the
species ML-MCTDHB expansion, where one replaces the expansion (\ref{spc_spf}) by an expansion in
terms of Hartree products made of the SPFs $|\phi^{3;\kappa}_i\rangle$, requires $M^S+SMm^N+SNmn$
coefficients, where $M$ is the number of species SPFs of each species. For a direct
standard MCTDHBs expansion, one needs $\binom{N+m-1}{m-1}^S+Smn$ coefficients. Finally,
the species ML-MCTDHB ansatz consists of $M^S+SM\binom{N+m-1}{m-1}+Smn$
coefficients. Table \ref{tab_scaling} lists the memory consumption of the
different methods for $S=2$ species, $n=250$ spatial
grid points, $m=4$ particle SPFs and various numbers of bosons in each species, $N$, and species SPFs, $M$. If the
necessary number of species SPFs is not too large with respect to the total
number of bosonic configurations, $\binom{N+m-1}{m-1}$, ML-MCTDHB clearly
requires less memory than all the other methods. In the case of the two
species example and $N=40$, $M=4$, the ML-MCTDHB wave function ansatz
consists of three orders of magnitude less coefficients than the corresponding
MCTDH-BB ansatz.
For large numbers of species SPFs, however, a MCTDH-BB expansion becomes
preferable with respect to
the memory consumption, which can be seen from the $N=4$ and
$M=16,35$ example in table \ref{tab_scaling}.

Obviously, the number of species and particle SPFs,
$i.e.$ (M,m), needed for a converged simulation does strongly depend on the details of the system:
The more correlations and entanglement are present in the system the more of these basis functions
are required. In view of this, table \ref{tab_scaling} serves as an exemplary overview over the (M,m)
parameter space being accessible by ML-MCTHB but not or hardly by other methods. For instance,
we can interpret the last row of table \ref{tab_scaling} as that ML-MCTDHB can simulate a particular
parameter regime of a two-species bosonic mixture containing forty bosons of each species, where
the simulation is converged with the
given (M,m), while the huge number of expansion coefficients prevent other methods to reach any
parameter regime except where a mean-field approximation is valid.

\begin{table}[htb]

 \centering

\begin{tabular}{|c|c|c|c|c|}

\hline

 (N, M) & MCTDH & ML-MCTDH & MCTDH-BB & ML-MCTDHB\\

\hline\hline

 (4,2) & $7.4\!*\!10^4$ / 34.3 & $9.0\!*\!10^3$ / 4.2  & $3.2\!*\!10^3$ / 1.5 &

$2.1\!*\!10^3$ / 1.0\\

\hline

 (4,6) & $7.4\!*\!10^4$ / 30.0 & $1.1\!*\!10^4$ / 4.5  & $3.2\!*\!10^3$ / 1.3 &

$2.5\!*\!10^3$ / 1.0\\

\hline

 (4,14) & $7.4\!*\!10^4$ / 23.2 & $1.5\!*\!10^4$ / 4.8  & $3.2\!*\!10^3$ / 1.02

&

$3.2\!*\!10^3$ / 1.0\\

\hline

(4,16) & $7.4\!*\!10^4$ / 21.8 & $1.6\!*\!10^4$ / 4.9  & $3.2\!*\!10^3$ / 0.96 &

$3.4\!*\!10^3$ / 1.0\\

\hline

(4,35) & $7.4\!*\!10^4$ / 13.0 & $2.7\!*\!10^4$ / 4.8  & $3.2\!*\!10^3$ / 0.6 &

$5.7\!*\!10^3$ / 1.0\\

\hline

(40,1) & $1.5\!*\!10^{48}$ / $5.5\!*\!10^{43}$ & $2.4\!*\!10^{24}$ /

$9.1\!*\!10^{19}$  & $1.5\!*\!10^8$ / $5.7\!*\!10^3$ &

$2.7\!*\!10^4$ / 1.0\\

\hline

(40,2) & $1.5\!*\!10^{48}$ / $2.8\!*\!10^{43}$ & $4.8\!*\!10^{24}$ /

$9.4\!*\!10^{19}$  & $1.5\!*\!10^8$ / $3.0\!*\!10^3$ &

$5.1\!*\!10^4$ / 1.0\\

\hline

(40,3) & $1.5\!*\!10^{48}$ / $1.9\!*\!10^{43}$ & $7.3\!*\!10^{24}$ /

$9.5\!*\!10^{19}$  & $1.5\!*\!10^8$ / $2.0\!*\!10^3$ &

$7.6\!*\!10^4$ / 1.0\\

\hline

(40,4) & $1.5\!*\!10^{48}$ / $1.5\!*\!10^{43}$ & $9.7\!*\!10^{24}$ /

$9.6\!*\!10^{19}$  & $1.5\!*\!10^8$ / $1.5\!*\!10^3$ &

$1.0\!*\!10^5$ / 1.0\\

\hline

\end{tabular}

\caption{The scaling of the methods MCTDH, ML-MCTDH, MCTDH-BB and ML-MCTDHB is
compared for the case of $S=2$ species, $m=4$ particle SPFs and $n=250$
spatial grid points. The number of bosons per species, $N$, and (for the
multi-layer methods) the number of species SPFs, $M$, are varied. Each table
entry contains the number of coefficients needed for the wave function expansion of
the respective method and its ratio with respect to the number of ML-MCTDHB
coefficients.}

\label{tab_scaling}

\end{table}

\subsection{Symmetry Conservation}\label{subsec_symm}
Let us focus next on the symmetry conservation property of ML-MCTDHB. The discussion in this section can
be extended to the complete MCTDH family, including MCTDH, ML-MCTDH and MCTDHB(F). It is already known that
the equations of motion of the MCTDH family preserve the total energy and the normalization of the wave
function \cite{BJWM00}. Here we show that the equations also preserve a certain class of symmetries, given
that the initial wave function and SPFs on each level possess the symmetry for the SPFs on each level. To illustrate this idea, we
focus on the particle ML-MCTDHB and assume that the Hamiltonian $\hat H$ has some symmetry group $\mathcal{G}$
and let $\hat G$ be an element of the unitary representation of $\mathcal{G}$ in the many-body Hilbert space,
i.e. $[\hat G,\hat H]=0$. We hereby only consider symmetry operations which can be decomposed as
$\hat G=\bigotimes_i\hat g_i$, where $\hat g_i$ is an element of the unitary representation of
$\mathcal{G}$ in the Hilbert space of the $i$-th primitive DOF.
(Some of the $\hat g_i$-th may be the unit operator.) $\hat G$ can be further represented as a
product of operators acting on the $\kappa$-th logical DOF of the $\alpha$-th layer:
$\hat G=\bigotimes_\kappa \hat g^{\alpha;\kappa}$. Let us assume that initially the total wave function and all
SPFs are invariant under the respective symmetry operators, 
$i.e.$ $\hat G |\Psi(0)\rangle = e^{i\Theta}|\Psi(0)\rangle$ and 
$\hat g^{\alpha;\kappa}|\phi^{\alpha;\kappa}_j(0)\rangle=e^{i\theta^{\alpha;\kappa}_j}|\phi^{\alpha;\kappa}_j(0)\rangle$.
Then the Hamiltonian matrix $\langle \vec n|\hat H |\vec m\rangle$
in (\ref{eq_parml}) does not allow transitions from the permanent state $|\vec m\rangle$
with $\hat G |\vec m\rangle = e^{i\Theta'}|\vec m\rangle$ to $|\vec n\rangle$
with $\hat G |\vec n\rangle = e^{i\Theta''}|\vec n\rangle$ and
$\Theta' \mod 2\pi\neq\Theta'' \mod 2\pi$, when propagating the coefficients for
a small time step
$\Delta t$, i.e. $\langle \vec n|\hat H |\vec m\rangle$ vanishes for such permanent states.
Hence, we are left to show that the SPFs preserve their symmetry
up to $O(\Delta t^2)$ when propagating them for a small time step. We note
that the mean-field operator matrices of a Hamiltonian containing at most
$F$-body terms can be expressed by the $F$-body reduced density matrices and
one-body operators. Now the initially well defined many-body symmetry, 
$A_{\vec n}(t=0)\neq 0\Rightarrow \hat G |\vec n\rangle(t=0) = e^{i\Theta}|\vec n\rangle(t=0)$,
implies that the $F$-body reduced density matrix elements
$\rho^{F;\alpha;\kappa}_{i_1,...,i_F;j_1,...,j_F}$ can only be non-vanishing
if
$\sum_{r=1}^F\theta^{\alpha;\kappa}_{i_r}=\sum_{r=1}^F\theta^{\alpha;\kappa}_{j_r}$
up to integer multiples of $2\pi$.
This can be seen implicitly from the mean field
matrices (\ref{mf_pot_inter})
in the example of two-body interactions.
If we further group the initial SPFs $|\phi^{\alpha;\kappa}_j(0)\rangle$ into
classes of the same $\theta^{\alpha;\kappa}_j\mod 2\pi$, the inverses of the
one-body reduced
density matrices in (\ref{dmat_pml}) are initially block diagonal. With these
two observations,
one can directly show 
$\hat g^{\alpha;\kappa}\partial_t |\phi^{\alpha;\kappa}_j(t)\rangle=e^{i\theta^{\alpha;\kappa}_j}\partial_t|\phi^{\alpha;\kappa}_j(t)\rangle$
at $t=0$. The same arguments hold for
any later time steps as long as $\Delta t$ is small enough and the symmetry conservation
becomes exact for $\Delta t\rightarrow 0$.	

Following the proof, we conclude that if at $t=0$ the many-body wave function $|\Psi(0)\rangle$
and all the SPFs $|\phi^{\alpha;\kappa}_j(0)\rangle$ are invariant under $\hat G$ and $\hat g^{\alpha;\kappa}$,
respectively, they remain invariant for all times, i.e. $\hat G |\Psi(t)\rangle=e^{i\Theta}|\Psi(t)\rangle$ and
$\hat g^{\alpha;\kappa} |\phi_j(t)\rangle=e^{i\theta^{\alpha;\kappa}_j}|\phi^{\alpha;\kappa}_j(t)\rangle$ for
all $t$. Such an initial state preparation can be performed in ML-MCTDHB by choosing the initial SPFs as eigenstates
of the single particle symmetry operators and populating only number states with the same many-body symmetry. Please
note that this symmetry conservation property cannot be expected {\it a priori} due to the truncation of the many-body
Hilbert space and the complicated integro-differential equations for the SPFs and to the best of our knowledge this has
not been shown previously. Possibly this symmetry conservation can be employed for specifying the configuration 
selection in selected configuration MCTDH type methods (cf. \cite{MW03} and references therein).	

\section{Implementation}\label{sec_impl}
\subsection{Implementation of the Second Quantization Formalism}\label{subsec_ns}
  In this section we briefly comment on some technical details of the ML-MCTDHB algorithm, and we mainly
  focus on two issues: the permanent state sequencing and the second-quantization calculations, which
  manifest themselves as a major challenge in the numerical treatment of the bosonic systems.
  
  A permanent state of $|\vec{n}\rangle=|(n_1,n_2,...,n_M)\rangle$ can be naturally viewed as an array of M
  integers with the constraint $n_1+n_2+...+n_M=N$. Taking a single species bosonic system for instance, the
  coefficients of all $|\vec{n}\rangle$, $e.g.$, $A_{\vec{n}}$ are normally stored in a one-dimensional array,
  and the index $I$, $i.e.$ the position, of $A_{\vec{n}}$ in the array is given by an indexing function of
  $I(\vec{n})$. Due to the constraint of $n_1+n_2+...+n_M=N$, we make use of the so-called Combinadic
  numbers \cite{SAC10,COMBI64} for the indexing. The Combinadic numbers can be characterized as follows:
  firstly $\vec{n}$ is related to an integer $n_1*N^{M-1}+...+n_i*N^{M-i}+...+n_M$, then all $\vec{n}$ are sorted
  by the descending order of this integer, and $I(\vec{n})$ is defined as the sequence number of $\vec{n}$ in the
  sequence. For example, we have $I[(N,0,...,0)]=1$, $I[(N-1,1,0,...,0)]=2$ 
  and $I[(0,...,0,N)]=\binom{N+M-1}{M-1}$. In the ML-MCTDHB implementation the more important function is actually
  the inverse function of $I(\vec{n})$, named as $\vec{n}(I)$. We build a table $T_1$ of a $\binom{N+M-1}{M-1}\times M$
  matrix to represent $\vec{n}(I)$, of which the element of $(T_1)_{ij}$ stores the occupation number in the $j$-th orbital
  of the $i$-th permanent state in the descending order. This table only needs to be built once at the beginning of the
  calculation with an overall negligible CPU time.
  
  The second-quantization calculations mainly appear in the equations of motion for the species coefficients and also the
  mean-field operators of the particle layer. Here we demonstrate the strategy applied in ML-MCTDHB for such second-quantization
  calculations with the example of the equations of motion. The equations of motion for $A_{\vec{n}}$ can be summarized as 
  \begin{equation}\label{eq_looping}
   i\partial_tA_{\vec{n}} = \sum_{i,j}\langle\vec{n}|\hat{H}_{1b}|\vec{n}+\hat{i}-\hat{j}\rangle A_{\vec{n}+\hat{i}-\hat{j}}+\sum_{i_1,i_2,j_1,j_2}\langle\vec{n}|\hat{H}_{2b}|\vec{n}+\hat{i}_1+\hat{i}_2-\hat{j}_1-\hat{j}_2\rangle A_{\vec{n}+\hat{i}_1+\hat{i}_2-\hat{j}_1-\hat{j}_2}.
  \end{equation}
  The first term on the rhs of the equation is related to the one-body operator, and the second term is related to the two-body operator.
  Equation (\ref{eq_looping}) must be applied to all permanent states. To calculate the rhs of (\ref{eq_looping}), firstly we need to loop
  over all the permanent states of $\vec{n}$, and for each permanent state two more loops over the orbitals $i,j$ and $i_1,i_2,j_1,j_2$ are
  required for the one-body and two-body operators, which ends up with a loop of the total number $\binom{N+M-1}{M-1}*(M^2+M^4)$.
  In each loop unit, the function  $I(\vec{n}+\hat{i}-\hat{j})$ or $I(\vec{n}+\hat{i}_1+\hat{i}_2-\hat{j}_1-\hat{j}_2)$ must be calculated,
  and this leads to huge numerical costs. The second quantization calculation for the mean-field operators are done in the same way as the equations of motion.
  
  However, in the ML-MCTDHB implementation, we use another strategy which can not only reduce the total amount of loops but meanwhile avoid
  the calculation of the index function $I(\vec{n}+\hat{i}-\hat{j})$ or $I(\vec{n}+\hat{i}_1+\hat{i}_2-\hat{j}_1-\hat{j}_2)$. To illustrate this,
  we rewrite the equation (\ref{eq_looping}) as 
  \begin{align}\label{eq_loop2}
  \sum_{\vec{n}}(i\partial_tA_{\vec{n}})|\vec{n}\rangle &=\sum_{\vec{p}|N-1}\sum_{i,j}|\vec{p}+\hat{i}\rangle\langle\vec{p}+\hat{i}|\hat{H}_{1b}|\vec{p}+\hat{j}\rangle A_{\vec{p}+\hat{j}} \\\nonumber
      &+\sum_{\vec{q}|N-2}\sum_{i_1,i_2,j_1,j_2}|\vec{q}+\hat{i}_1+\hat{i}_2\rangle\langle\vec{q}+\hat{i}_1+\hat{i}_2|\hat{H}_{2b}|\vec{q}+\hat{j}_1+\hat{j}_2\rangle A_{\vec{q}+\hat{j}_1+\hat{j}_2}.
   \end{align}
  The equation (\ref{eq_loop2}) can be interpreted as follows: we generate a one-dimensional array of length $\binom{N+M-1}{M-1}$, which
  stores the value of $\partial_tA_{\vec{n}}$ for all the permanent states. To fill in the array, $i.e.$, to calculate the time derivative
  of all the permanent states, we start with the one-body operators and loop over the permanent states $\vec{p}|N-1$ for $N-1$  bosons instead
  of those for $N$ bosons and calculate $\langle\vec{p}+\hat{i}|\hat{H}_{1b}|\vec{p}+\hat{j}\rangle A_{\vec{p}+\hat{j}}$ to put the value in
  the position of $I(\vec p+\hat i)$. For the two-body operators, we loop over the permanent states $\vec{p}|N-2$ for $N-2$  bosons, calculating
  $\langle\vec{q}+\hat{i}_1+\hat{i}_2|\hat{H}_{2b}|\vec{q}+\hat{j}_1+\hat{j}_2\rangle A_{\vec{q}+\hat{j}_1+\hat{j}_2}$ and putting it in the
  position $I(\vec{q}+\hat{i}_1+\hat{i}_2)$ of the array. In this way the total number of loops reduces from $\binom{N+M-1}{M-1}*(M^2+M^4)$ 
  to $\binom{N+M-2}{M-2}*M^2+\binom{N+M-3}{M-3}*M^4$.
  
  The strategy of equation (\ref{eq_loop2}) can also help to avoid calculating the indexing function $I(\vec{n})$, as $I(\vec{n})$ is only
  an intermediate step linking two permanent states, which can be done by means of two pre-build tables, named $T_{1b}$ and $T_{2b}$, of
  $\binom{N+M-2}{M-2}*M$ and $\binom{N+M-3}{M-3}*M^2$ arrays, respectively. In $T_{1b}$, the element $(T_{1b})_{a,i}$ stores the index of
  the permanent state $I(\vec{p}+\hat{i})$, 
  where the integer $a$ is the Combinadic number of $|\vec{p}\rangle$. Similarly, the element $(T_{2b})_{a,i+(j-1)*M}$ stores the index of
  the permanent state $|\vec{p}+\hat{i}+\hat{j}\rangle$, with $a$ being the Combinadic number of $|\vec{p}\rangle$. These two tables take
  almost negligible CPU time in the initialization step of the calculation. With equation (\ref{eq_loop2}), the total loops amount is greatly
  reduced and simultaneously the time-consuming calculations of the index functions is avoided, which leads to a great reduction of the CPU
  time of the overall calculations.

  In this section we have presented how we face the major challenge of the ML-MCTDHB implementation, namely how to deal with the initialization
  and calculations within the second quantization formalism. We demonstrate that by introducing three indexing tables $T_1$, $T_{1b}$ and $T_{2b}$,
  we have an efficient solution to the difficulties in the second-quantization treatment, which both reduce the looping times and the calculation
  cost in each loop unit. Moreover, the $T_{1b}$ and $T_{2b}$ are particularly constructed for the one-body and two-body operators, and such a strategy
  also enables us to conveniently extend to F-particle interactions, simply by generating a table containing the mapping from permanent state
  $|\vec p\rangle$ to $|\vec p+\hat{i}_1+...+\hat{i}_F\rangle$.

  \subsection{ML-MCTDHB Program}\label{subsec_app}
  Having introduced the theory of ML-MCTDHB, we now turn to a brief introduction to our code. Our ML-MCTDHB code is based on the general ML-MCTDH implementation
  \cite{VHDM11} in its recursive formulation \cite{Manth08}. We have developed our code based on the multilayer machinery
  of the Heidelberg MCTDH85 package
  \cite{MCTDH_Heidelberg}, and equipped it with the machinery of dealing with bosonic ensembles.
  Since this ML-MCTDH
  implementation is capable of handing arbitrary ML-MCTDH wave function
  expansions, we have extended the scheme to species and particle ML-MCTDHB as well as mixtures of both. Hence, in principle an arbitrary number of
  bosonic species in arbitrary dimensions can be treated. 

  As introduced in the previous sections, the ML-MCTDHB scheme is divided into various separate functional bricks, such as the initialization
  of the tree structure with the permanent tables $T_1$, $T_{1b}$ and $T_{2b}$, the construction of the Hamiltonian, the mean-field operators
  and the density matrices of different layers, and the combination of all these components to the rhs of the equations of motion to perform
  the integration. Our ML-MCTDHB code is then designed in a systematic structure, consisting of various modules, which accommodate the requirements
  of different functional bricks of the method. The code can now carry out propagations in the real and imaginary time, which correspond to
  the calculation of the dynamics and the relaxation to the ground state, respectively. Moreover, improved relaxation \cite{MW03} is also
  implemented in our code, in which the non-top node coefficients are still propagated in imaginary time while
  the top-node coefficients are obtained by direct diagonalization. In the improved relaxation, the 
  imaginary time propagation of non-top node coefficients and the diagonalization of top node coefficients are done
  recursively. When convergence is reached, both the top-node and non-top node coefficients become constant in the 
  diagonalization and imaginary time propagation, respectively, which corresponds to a stationary state of the system,
  $i.e.$ an eigenstate. By carefully choosing the initial state of the improved relaxation, we can obtain different eigenstates
  of the system, which lie in the truncated Hilbert space given by the ML-MCTDHB Ansatz.

  At the moment, our ML-MCTDHB scheme can deal
  with arbitrary two-body interactions - an extension to three- or four-body
  interactions would be straightforward. We adopt the product form of 
  the interactions, as introduced in section \ref{subsec_species}, and this allows a recursive
  construction of the Hamiltonian expressed and also the mean-field operators on different layers.
  Ultra-cold bosonic atoms constitute an important class of systems, which can be attacked by the ML-MCTDHB theory.
  In this particular context, the contact interaction turns out to be very relevant. Approximating the delta
  interaction potential by a narrow Gaussian, the POTFIT machinery can be applied
  for obtaining the desired product form (cf. \cite{ZMS08}). For
  one-dimensional settings, we have also developed an exact implementation
  of
  the contact interaction $V(x_1-x_2)=g\,\delta(x_1-x_2)$ {\it without} reshaping
  it into the product form in order to liberate
  the simulations from the artificial length scale induced by the Gaussian
  approximation of the delta function. Moreover, the direct implementation of the contact potential has speeded
  up the simulation by one order of magnitude in comparison to a product representation of the same accuracy.
  The main difference in the derivation of
  the equations of motion lies in the fact that one cannot separate the mean-field
  operators into a scalar matrix and a operator valued factor anymore,
  which breaks the recursive multi-layer formulation (cf. the appendix). For the explicit equations of motion
  with the efficient treatment of the contact interaction we refer the reader to \cite{mlb_3spc}.

  Depending on the concrete problem, we employ an appropriate discrete variable
  representation (DVR) of the SPFs \cite{BJWM00,DVR}, such as the (radial) harmonic oscillator DVR, the sine DVR,
  the exponential DVR and the Legendre DVR, which have been implemented in our code. So in the case of the exact
  implementation
  of the delta interaction, the grid point spacing defines the smallest length
  scale, thus an ultraviolet cutoff.

  For integrating the equations of motion, we employ either ZVODE, a
  variable-coefficient ordinary differential equation solver with
  fixed-leading-coefficient implementation \cite{ZVODE}, or DOPRI, an explicit
  Runge-Kutta method of order five with step size control and dense output
  \cite{DOPRI}. 

  We can also use the ML-MCTDHB scheme to calculate the expectation
  values of the combination of one-body and two-body operators. For instance, we can calculate the total
  energy or separately the kinetic energy and potential energy evolution, the one-body density and the two-body density matrices.
  Especially, the two-body density matrices are already calculated during the propagation as a building brick for the mean-field
  operators, so that such quantities can be obtained for free in terms of computational time (cf. also \cite{MCHBm}).

\section{Applications}\label{sec_exampl}

In the following, we focus on the tunneling dynamics of ultra-cold bosonic atoms in a one-dimensional double
well trap made of a harmonic trap
superimposed with a Gaussian at the trap centre, i.e. 
$V_{trap}(x)=x^2/2+h/\sqrt{2\pi s^2}*\exp(-x^2/2s^2)$ in harmonic oscillator units
$\hbar=m=\omega=1$. Firstly, we will compare the single species tunneling results of our ML-MCTDHB implementation with the results of the implementations of MCTDH \cite{MCTDH_Heidelberg}.
Then we apply species ML-MCTDHB to simulate the tunneling of a bosonic
mixture and compare the results to the ML-MCTDH simulations \cite{VHDM11}. Applications of particle ML-MCTDHB will be presented
somewhere else.

In both examples, we prepare the initial state by
modifying the trapping potential such that it becomes energetically favorable
for the bosons to be in the left well. The bosonic ensemble is then relaxed to the many-body ground state by propagating the equations
of motion in imaginary time. The resulting many-body state is finally propagated
in real time in the original double well trap.

\subsection{Single-Species Tunneling}
As a first application, we simulate the tunneling dynamics of a single-species
bosonic ensemble in
the double well. The tunneling dynamics of a bosonic ensemble in a
one-dimensional double well has been extensively 
studied, for both microscopic and macroscopic systems, with boson
numbers ranging from two to
the order of $10^6$ and even more
\cite{dw-gp1,dw-gp2,dw-bh,SSAC09,ZMS08,dw-mbh}. Experiments on the double-well
tunneling
have also been carried out \cite{dw_exp1,dw_exp2,dw_exp3}. Various theoretical
approaches
have been employed, for instance, the Gross-Pitaevskii
equation \cite{dw-gp1,dw-gp2}, the Bose-Hubbard model \cite{dw-bh,dw-mbh}
 as well as ab-initio methods: MCTDHB has been applied to the many-body system
\cite{SSAC09}, and  
calculations via MCTDH have been carried out for few-body systems \cite{ZMS08}.
The simulations based on the 
single-band approximation predict that the tunneling is suppressed for
interaction strengths above
some critical value, while the extended model predicts the weakening of such a
suppression in the strong
interaction regime, where higher bands effects cannot be neglected. In this way
the double well potential
manifests itself as a proper test bed for the higher band effects. In this
section we will simulate the 
double well tunneling with ML-MCTDHB, and perform a detailed analysis to
resolve
the higher bands effects.

\begin{figure}
\centering
\includegraphics[width=0.95\textwidth]{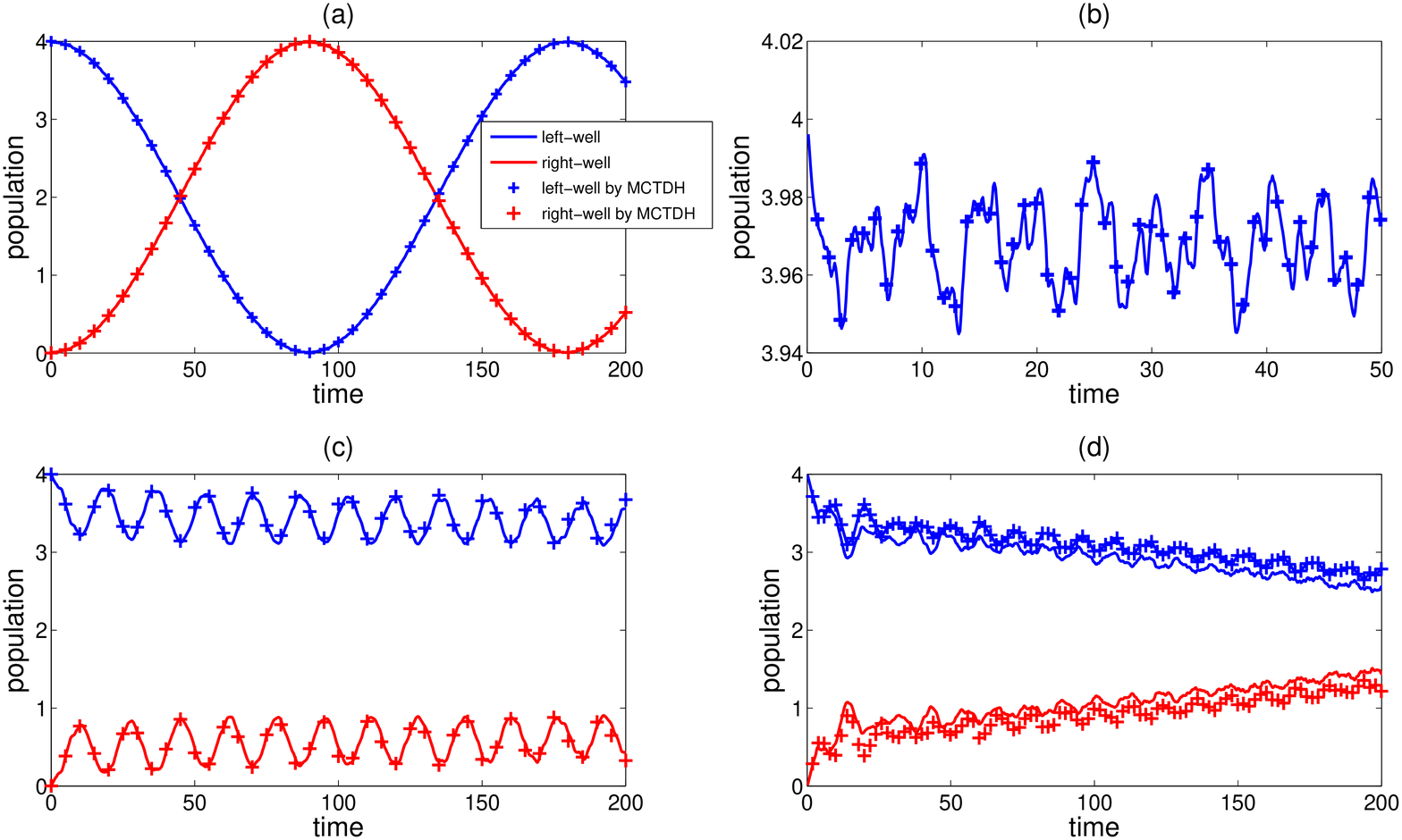}
\caption{The population oscillation of four bosons in the left and right well,
with (a) g=0.0, (b) g=0.5, (c) g=2.0, and
(d) g=4.0. Lines: ML-MCTDHB results. Crosses: MCTDH results. Figures (a)-(c): Six particle SPFs;
 figure (d): Ten particle SPFs. Particularly, figure (b) shows the behaviour
 for a shorter time interval to highlight the agreement between the two methods concerning the fast oscillation process.
The difference between the two methods in figure (d) is attributed to the different implementations 
 of the contact interaction in the two methods (cf. main text).}
\label{tunneling-4b}
\end{figure}

Here we present the tunneling dynamics of four and ten bosons
in a double well potential. A sin-DVR is employed, which
intrinsically introduces hard-wall boundaries at both
ends of the potential (cf. appendix of \cite{BJWM00}). Firstly we show the
simulation of the four-boson tunneling,
in comparison with the MCTDH simulation using the Heidelberg MCTDH package
\cite{MCTDH_Heidelberg}. To have
a direct comparison between ML-MCTDHB and MCTDH results, we adopt in both
simulations a narrow Gaussian
interaction to model the contact interaction, and the interaction is written as
$V(x_1-x_2)=g(2\pi\sigma^2)^{-1/2}\,\exp(-(x_1-x_2)^2/(2\sigma^2))$, with
$\sigma=0.05$.
Next we extend the simulation to the ten-boson case, and the contact interaction
is modeled
by the exact delta function. The simulation of the tunneling of ten bosons
becomes
impractical with MCTDH, and we only perform the simulation with ML-MCTDHB.

Figure \ref{tunneling-4b} summarizes the 
population evolution of 4 bosons in the double well for different interaction
strengths. When the interaction
strength is zero, the system undergoes Rabi oscillations, as shown in figure
\ref{tunneling-4b}(a). As 
the interaction strength increases to 0.5, the oscillations almost vanish
on a relative long time scale, which
indicates the delayed tunneling behaviour. When the interaction strength
increases to 2.0, as shown in figure
\ref{tunneling-4b}(c), the amplitude of the population oscillations is
increased
 from less than 0.06 in delayed tunneling (figure \ref{tunneling-4b}(b)) to
around 1.0, 
 which is referred to as enhanced tunneling. As the interaction increases 
even further to 4.0, the quasistationary state is approached during tunneling,
where the populations of the left and the right well approach the value of two 
with only small fluctuations, as shown in
figure \ref{tunneling-4b}(d). Summarizing, figure \ref{tunneling-4b}
illustrates the tunneling transition from Rabi oscillations through delayed
tunneling
and enhanced tunneling to the quasistationary state as the interaction strength
increases from zero to the strong interaction regime. The ML-MCTDHB results show
a
very good agreement with the MCTDH calculations. Only for the interaction
strength $g=4.0$, deviations occur, 
which can be explained by the different implementations of the POTFIT
algorithm in MCTDH and ML-MCTDHB. Nevertheless, we still observe qualitatively
the same behaviour
of the emergence of the quasistationary state in both simulations. Moreover, more
orbitals are needed to achieve
good convergence in the strong interaction case of $g=4.0$, and we supply ten
orbitals
to this case, where good convergence can be deduced from the natural populations
discussed in the following.

\begin{figure}
\centering
\includegraphics[width=0.95\textwidth]{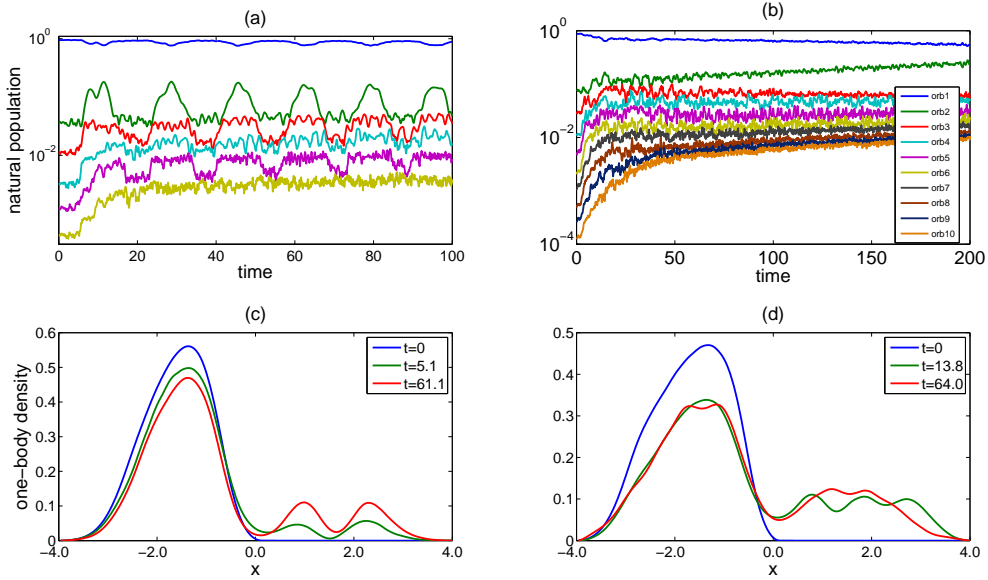}
\caption{The natural populations and one-body densities of the system of four
bosons in the double well at different time instants, with (a) and (c) for g=2.0,
as well as (b) and (d) for g=4.0.} \label{density-4b}
\end{figure}

 The natural
populations can confirm the convergence of the calculation and also manifest 
themselves as a measure of the fragmentation of the system \cite{fragmentation}, which is defined as the 
depletion of the population of the highest occupied natural orbital from unity. To uncover the
fundamental effect giving rise to
the enhanced tunneling and quasistationary state, we plot the natural populations
and
 one-body density profiles of the four-boson ensemble at different time instants
during the
tunneling process in figures \ref{density-4b}.
Figure \ref{density-4b}(a) and (c) show the natural populations and one-body
densities for $g=2.0$. In figure \ref{density-4b}(a), firstly we see the lowest
natural population saturates
to a value less than $10^{-3}$, which confirms the convergence of the
simulation. In figure
\ref{density-4b}(c), 
we observe that the profile in the left well remains as a Gaussian packet,
while 
the profile in the right well presents a two-hump structure. The two-hump
profile is a signature of the occupation
of the first excited state
in the right well, and this indicates that the enhanced tunneling is due to the
higher band occupation,
$i.e.$ the interband tunneling. In the 2-boson ensemble \cite{ZMS08} 
the enhanced tunneling only takes place in the fermionization regime, $i.e.$
for the 
interaction strength approaching infinity,
while as the number of bosons increases, it becomes easier to excite higher
bands, and the enhanced
tunneling arises even in an interaction regime far below the
fermionization limit. 

The natural populations for $g=4.0$,
as shown in figure \ref{density-4b}(b) show a good convergence of the
simulation with
the lowest natural population saturating well below 1 percent, and at $g=4.0$
more natural
orbitals contribute to the tunneling process, which suggests that fragmentation
of the system and the presence of multiple tunneling channels in the dynamics. Figure \ref{density-4b}(d) shows the one-body densities
at different times for the
interaction strength $g=4.0$, where the
quasistationary state dominates the tunneling. 
During the tunneling, the one-body density profile presents multiple oscillations in
both the left and right wells,
which indicates multiple higher-band excitations in the tunneling process. In
the exact quantum dynamics study
of the double well system \cite{SSAC09}, the quasistationary state is explained by
the quick loss of coherence of the system, and the multiple excitation of higher energetic levels in the left and
right well suggests that the
tunneling process involves a large number of higher band number states, and in
consequence multiple tunneling channels.
The dephasing between these tunneling channels can be a source of the
loss of coherence of the bosons in the two wells.

\begin{figure}
\centering
\includegraphics[width=0.95\textwidth]{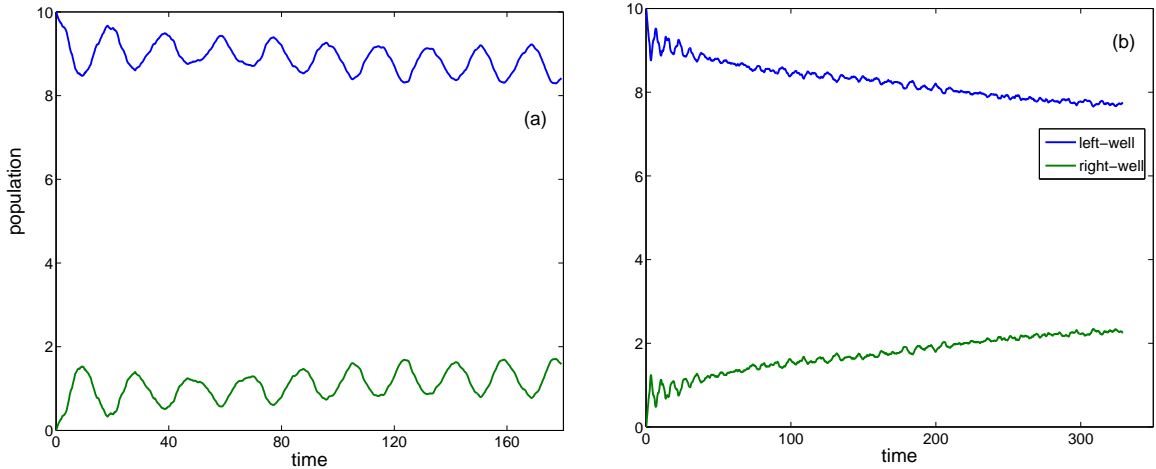}
\caption{The population oscillation of ten bosons in the double well, with (a)
g=0.5, (b) g=1.0. 
Enhanced tunneling is obtained in (a), where the amplitude of the population is
around 2. A slow
evolution to the quasistationary state is obtained in (b).} \label{tunneling-10b}
\end{figure}

\begin{figure}
\centering
\includegraphics[width=0.95\textwidth]{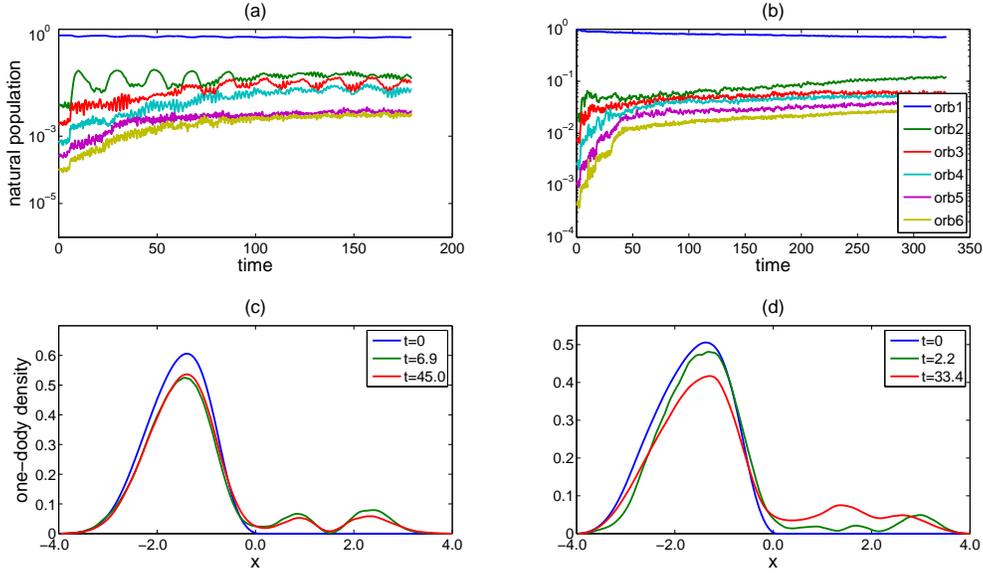}
\caption{The natural population and one-body densities of ten bosons in the
double well at different
time, with (a) and (c) g=2.0, as well as (b) and (d) g=4.0.} \label{density-10b}
\end{figure}

Figure \ref{tunneling-10b} shows the tunneling evolution of a system of ten
bosons in the double well with varying interaction strength. We focus on the
enhanced tunneling and quasistationary state for a 
 sufficiently large interaction strength. Enhanced tunneling is observed
with an interaction strength as weak as $g=0.5$,
as shown in figure \ref{tunneling-10b}, and at $g=1.0$ the system slowly
evolves
to the quasistationary state. Figures \ref{density-10b}(a) to (d) show the natural
populations and
 one-body densities at different time instants during
the tunneling process of figures \ref{tunneling-10b}(a) and (b), respectively.
The fact that the lowest natural population saturates to relatively small values
of the order of 0.1\% and 1\% in figures \ref{density-10b} (a) and (b), respectively,
illustrates the well-controlled behavior of the convergence of the simulation.
In figure \ref{density-10b}(c), the 
two-hump profile in the right well indicates that the enhanced tunneling is
again a result of interband tunneling,
and the multi-mode oscillatory structure in figure \ref{density-10b}(d) suggests that
multiple tunneling channels are involved and
this can lead to the decoherence between the two wells, and consequently to the
appearance of the quasistationary state.

To summarize, in this section we presented the tunneling dynamics of single species bosons in a one-dimensional double well potential.
We supply the cross check with results obtained by the Heidelberg MCTDH, which indicates the stable performance of ML-MCTDHB,
and also the check of convergence by the natural populations. Further, we also demonstrate the ability of the method for various and extended investigations,
via different analysis routines, such as the population evolution and the one-body density evolution for larger systems.
\subsection{Mixture Tunneling}

Let us now consider the tunneling dynamics of two bosonic species, called the A and B species
which are loaded in the left well of the double well trap. All
species shall have the same mass, which is set to one,
and shall experience the same double well potential $V_{trap}(x)$. The
intra-species interaction strengths $g_\sigma$ of the contact interaction
potentials $V_\sigma(x)=g_\sigma \delta(x)$ $(\sigma=A,B)$, however, are assumed to be
different for different species $\sigma$.
Furthermore, the inter-species
interaction is also modelled by a pseudo-potential:
$V_{AB}(x_A-x_{B})=g_{AB}\delta(x_A-x_B)$. All delta potentials are implemented
numerically exactly as explained in \cite{mlb_3spc}. The choice $h=3$ and $s=0.2$ 
provides us with three bands below the barrier energy with two single particle 
eigenstates each. The energetic separation of the lowest band to the first one amounts to 1.63,
while the level spacing of the lowest band equals 0.23 resulting in a Rabi-tunneling period of 27. For preparing the
initial state of the mixture, we modify the double well trap
by letting $V_{trap}(x)=20.0$ for $x>0$. 

We consider
a binary mixture made of 2 A and 2 B bosons. With $g_A=0.3$, $g_B=0.5\,g_A$ and $g_{AB}=0.1\,g_A$. Due to the not too different
intra-species
interaction strengths, we provide for each species the same number of species
SPFs,
$M_\sigma\equiv M=4$, and particle SPFs, $m_\sigma\equiv m=3$. In the following, we compare simulations done
with ML-MCTDHB and ML-MCTDH \cite{VHDM11}. Although the initial state for the relaxation run
is - from a mathematical perspective - perfectly symmetric with respect to
particle exchange within each species, one has to pay attention to the initially
unoccupied species SPFs in ML-MCTDH. These have to be symmetrized 'by hand'
because otherwise the ML-MCTDH propagation will not preserve the exchange
symmetry within each species. 

MCTDH and its derivative methods are proven to preserve both the norm and the
total energy exactly (cf. appendix of \cite{BJWM00}).
Using the ZVODE integrator with an absolute and relative tolerance of
$10^{-10}$ and integrating $100$ harmonic oscillator time units, the norm of the
total wave function deviates from unity by $10^{-7}$ and the total
energy is conserved up to $10^{-6}$ for both ML-MCTDH and ML-MCTDHB. 

\begin{figure}[h]
 \centering
 \includegraphics[width=0.65\textwidth]{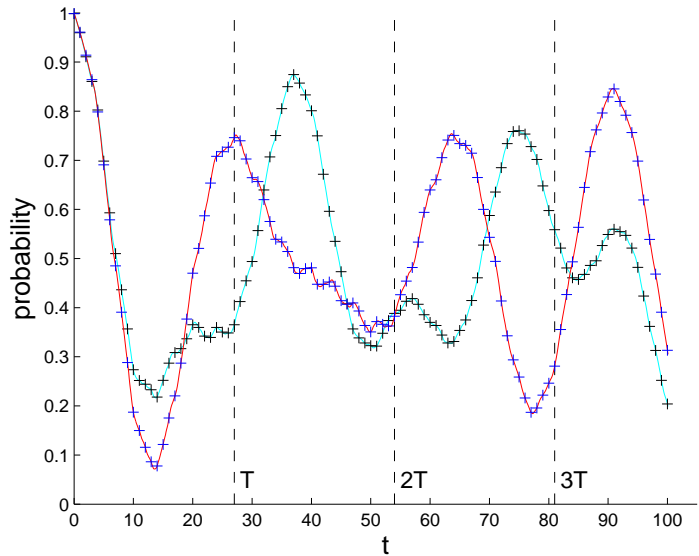}
 \caption{2 A and 2 B bosons initially loaded in the left well
of a double well trap. Blue (red) line: Probability for finding an
A (B) boson in the left well versus time. Line: ML-MCTDHB
results. Crosses via ML-MCTDH.
Parameters: $g_A=0.3$, $g_B=0.5\,g_A$ and $g_{AB}=0.1\,g_A$.
$M=4$ and $m=3$ SPFs for both calculations. The dashed vertical lines: the first three Rabi-tunneling
periods of a single particle.}
\label{fig_2x2_tunnel_prob}
\end{figure}

\begin{figure}[h]
 \centering
 \includegraphics[width=0.65\textwidth]{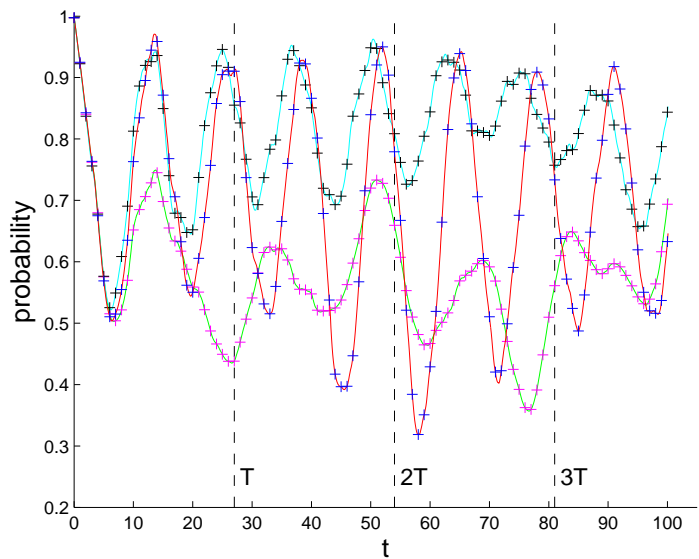}
 \caption{Evolution of several joint probabilities for the same setup
as in figure \ref{fig_2x2_tunnel_prob}: The blue (red) solid line:
probability of finding both A (B) bosons in the same well; The green solid line: The probability of
detecting an A and an B boson in the same well. Solid lines: ML-MCTDHB. Crosses: ML-MCTDH. Dashed vertical lines: The first three Rabi-tunneling periods.}
\label{fig_2x2_tunnel_jointprob}
\end{figure}

Before discussing the convergence of the simulations with respect to the number
of
particle and species SPFs, we first summarize the results for the different one- and
two-body observables. Figure \ref{fig_2x2_tunnel_prob} shows the time evolution
of the probability for finding an A (B) boson in the left well. One clearly
sees that the tunneling period of the A bosons is enlarged in comparison to the
Rabi-tunneling period. In contrast to this, the probability evolution of
finding a B boson in the left well qualitatively resembles a first Rabi-cycle
but afterwards also features a delay. This observation is quite
plausible: Since both the $g_B$ and $g_{AB}$ are smaller than $g_A$, one expects that the B bosons require a longer interaction time in
order to show an interaction-induced effect. The impact of the different
interaction strengths can also be seen in figure \ref{fig_2x2_tunnel_jointprob}:
While the probability for finding two A bosons in the same well is well above
0.5 for most of the propagation time, showing a binding tendency, the B bosons tend to stay in the same well
less likely. On the contrary, the probability for finding an A and a B
boson in the same well fluctuates around 0.5 indicating that the bosons of each
species tunnel independently.

\begin{figure}[h]
 \centering
 \includegraphics[width=0.6\textwidth]{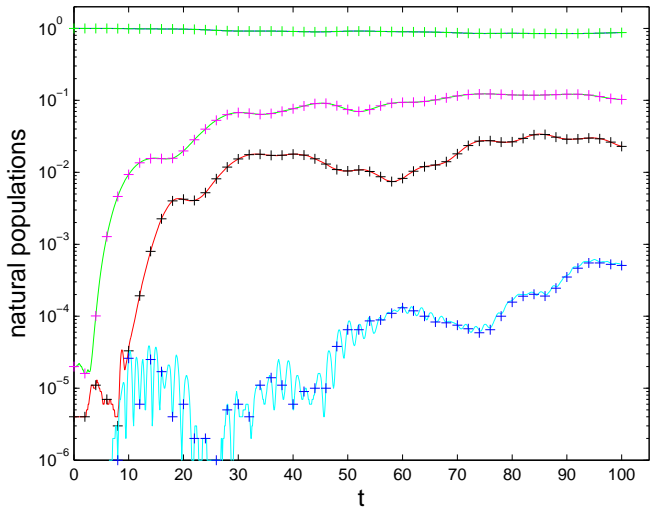}
 \caption{The natural populations, i.e. eigenvalues of the reduced density
matrix corresponding to the whole species A (or equivalently B) are plotted
versus time for the same tunneling scenario as in figure
\ref{fig_2x2_tunnel_prob}. The solid lines refer to the ML-MCTDHB and the
crosses
to the corresponding ML-MCTDH calculation.}         
\label{fig_2x2_natpop_spec}
\end{figure}

\begin{figure}[h]
\centering
\subfigure[]{
 \centering
 \includegraphics[width=0.55\textwidth]{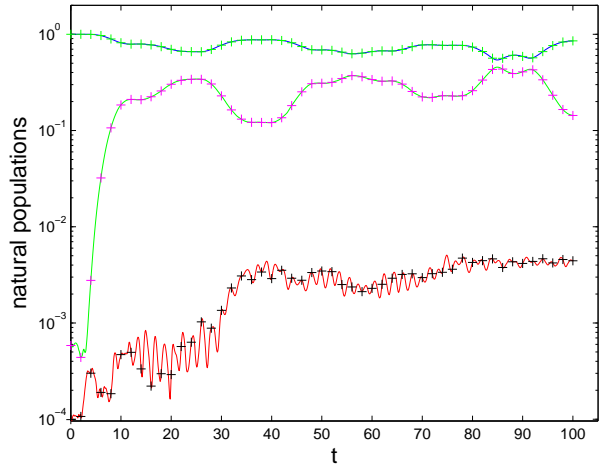}
 \label{sfig_2x2_natpop_partA}
}
\subfigure[]{
 \centering
 \includegraphics[width=0.55\textwidth]{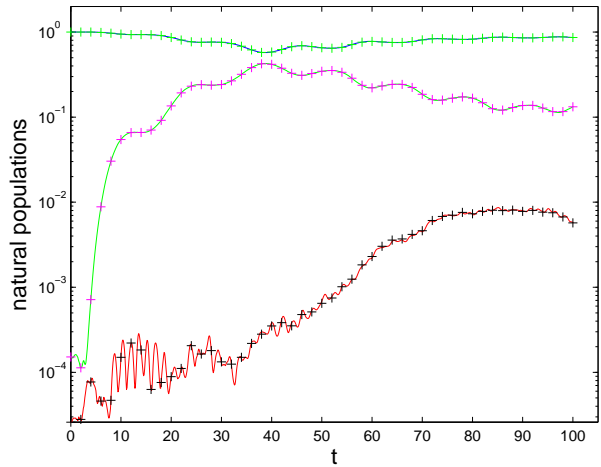}
 \label{sfig_2x2_natpop_partB}
}
 \caption{The dynamics of the natural populations of the reduced density matrix
corresponding to an A and a B boson are shown in
(a) and (b), respectively. All
parameters are chosen as in figure
\ref{fig_2x2_tunnel_prob}. The solid lines: ML-MCTDHB; the
crosses: ML-MCTDH calculation.}         
\label{fig_2x2_natpop_part}
\end{figure}

For judging the convergence of the simulations, we present the natural populations
for different subsystems. Figure \ref{fig_2x2_natpop_spec} shows the natural
populations
corresponding to the reduced density matrix of the whole species A (or B). One
clearly sees that after about
25 time units three species states contribute to the total wave function with
weights of the order of 89\%, 10\% and 1\%. The fourth state contributes so
little that it could be neglected without affecting the results. 
The natural
populations of the reduced density matrix corresponding to an A and a B boson,
respectively, are plotted in figure \ref{fig_2x2_natpop_part}. Here again,
we notice that the lowest natural population stays well below 1\%, meaning that
its natural orbital, the corresponding eigenstate of the respective reduced
one-body density matrix, has only marginal influence on the result.
Furthermore, we observe that two natural orbitals contribute with almost equal
weight during certain time intervals. Hence, a mean-field approximation would be
improper, which was to be expected for such a few-body system. In terms of the numerical correctness of
our implementation, we note that the ML-MCTDHB results excellently agree with the simulations performed with
ML-MCTDH.

In this section we have demonstrated the correctness of the implementation of ML-MCTDHB in comparison with MCTDH and ML-MCTDH,
with an excellent agreement being observed. It is worth pointing out that ML-MCTDHB is not only more efficient, but allows us to treat more complicated systems with more particles, more species and for stronger correlations. For an application of ML-MCTDHB to a more involved tunneling scenario of a bosonic mixture, we refer the reader to ref \cite{mlb_3spc}.

\section{Conclusions and Outlook}\label{sec_concl}

The ab-initio methods MCTDHB and ML-MCTDH for the investigation of many-particle quantum systems possess different emphases and foci: While
MCTDHB aims at employing the bosonic particle exchange symmetry for obtaining a
better performance, ML-MCTDH focuses on how to obtain a more compact ansatz for
the many-body wave function by employing physical knowledge about correlations within and between
\textit{subsystems}.
In general however, the high flexibility of the ML-MCTDH wave function
ansatz is incompatible with the generic correlations due to the
bosonic exchange symmetry in a MCTDHB wave function expansion.

In this work, we have shown that one can benefit from the advantages of both the
MCTDHB and the ML-MCTDH concept if one restricts oneself to two - quite natural
- classes of wave function expansion schemes (and any combination thereof): (i)
In species ML-MCTDHB, the total wave function of a bosonic multi-component
system firstly is expanded in Hartree products made of states with each of them
corresponding to a state of a whole species. Then each of these species states
is expanded in permanents like in MCTDHB. (ii) In particle ML-MCTDHB, a single
component bosonic system is considered with the bosons living in two or three
dimensions and/or having internal degrees of freedom. There, the total wave
function is expanded in terms of permanents and a ML-MCTDH expansion is applied
to all the orbitals underlying those permanents. Summarizing, the bosonic
exchange symmetry is exactly and efficiently taken into account in ML-MCTDHB as
any state of indistinguishable bosons is expanded in permanents. The
multi-layer concept is then employed for obtaining an optimized wave function
ansatz guided by the correlations between different species (intra- versus
inter-species correlations) or between different spatial directions (e.g. in
quasi two- or quasi three-dimensional systems) or internal degrees of freedom.
By comparing with other methods of the MCTDH family, we have demonstrated the beneficial scaling of ML-MCTDHB.
Like in all MCTDH type methods, the convergence of a simulation with respect to
the number of provided states in the wave function ansatz can be judged by the
eigenvalue distributions of the reduced density matrices corresponding to
different subsystems. 

We have implemented ML-MCTDHB on the basis of the ML-MCTDH scheme \cite{VHDM11,MCTDH_Heidelberg}
in such a general way that, in principle, we can deal with an arbitrary number
of species in arbitrary dimensions with different types of interactions, such as
the contact as well as the dipolar interactions
- only limited by the number of states needed
for a converged simulation, which depends on the system details, of course.
Furthermore, the scheme can in principle also deal with hybrid systems such as a
single ion coupled to an environment of indistinguishable bosons like liquid $^4\text{He}$ as long as all
the interactions can be efficiently brought into the POTFIT product form.

Finally, we note that both conceptually and practically it is
relatively straightforward to generalize the presented method to fermionic
systems, i.e. to ML-MCTDHF, or to mixed bosonic-fermionic systems, i.e. to
ML-MCTDHBF: Similar to equation (\ref{spc_spf}), one could
start with an appropriate ML-MCTDH expansion in which all indistinguishable
particles of one kind are grouped into a species node, and expand the species SPFs
with respect to the permanent states and slater determinants for bosonic and fermionic
species, respectively. This approach is in particular compatible with our ML-MCTDHB method
and its implementation, and would lead to a highly efficient algorithm for simulating the
most general composite systems consisting of subsystems with indistinguishable constituents.

\begin{acknowledgments}
 The authors would like to thank Hans-Dieter Meyer and Jan Stockhofe for
fruitful discussions on MCTDH methods and symmetry conservation.
Particularly, the authors would like to thank Jan Stockhofe for the DVR
implementation of the ML-MCTDHB
code. We also thank Johannes Schurer and Valentin Bolsinger for carefully reading of
the manuscript. S.K. gratefully acknowledges a scholarship of the Studienstiftung
des deutschen Volkes. L.C. and P.S. gratefully acknowledge funding by the
Deutsche Forschungsgemeinschaft in the framework of the SFB 925 ``Light induced
dynamics and control of correlated quantum systems''.
\end{acknowledgments}

\appendix*
\section{Species Multi-layer MCTDHB Equations of Motion without Product Form of the Hamiltonian} \label{spc_mctdhb}

  Here we provide the explicit equations of motion without using the product form of the interactions. We first introduce the Hamiltonian in the second quantization picture, as 
  \begin{align} \label{ham_par}
  H&=\sum_{\kappa}\sum_{i,j}(H^{3;\kappa}_0)_{ij}\hat{a}^\dagger _{\kappa,i}\hat{a}_{\kappa,j} \\\nonumber
     &+\frac{1}{2}\sum_\kappa\sum_{i_1,i_2,j_1,j_2}(V^{3;\kappa})_{i_1i_2j_1j_2}\hat{a}^\dagger _{\kappa,i_1}\hat{a}^\dagger _{\kappa,i_2}\hat{a}_{\kappa,j_1}\hat{a}_{\kappa,j_2} \\\nonumber
     &+\sum_{\kappa_1<\kappa_2}\sum_{i_1,i_2,j_1,j_2}(W^{3;\kappa_1,\kappa_2})_{i_1i_2j_1j_2}\hat{a}^\dagger _{\kappa_1,i_1}\hat{a}^\dagger _{\kappa_2,i_2}\hat{a}_{\kappa_1,j_1}\hat{a}_{\kappa_2,j_2}.   
  \end{align}
 $\hat{a}_{\kappa,i}$ ($\hat{a}^\dagger _{\kappa,i}$) refers to the operator annihilating (creating) a
 $\kappa$ boson in the SPF state $|\phi^{3;\kappa}_i\rangle$. The coefficients in the Hamiltonian terms
 defined by the standard second quantization are
 $(H^{3;\kappa}_0)_{ij}=\langle \phi^{3;\kappa}_i|\hat H^0_\kappa|\phi^{3;\kappa}_j\rangle$, 
 $(V^{3;\kappa})_{i_1i_2j_1j_2} =\langle \phi^{3;\kappa}_{i_1}|\langle \phi^{3;\kappa}_{i_2}
 |\hat{V}|\phi^{3;\kappa}_{j_1}\rangle|\phi^{3;\kappa}_{j_2}\rangle$
 and $(W^{3;\kappa_1,\kappa_2})_{i_1i_2j_1j_2}=\langle \phi^{3;\kappa_1}_{i_1}|\langle \phi^{3;\kappa_2}_{i_2}
 |\hat{W}^{\kappa_1,\kappa_2}|\phi^{3;\kappa_1}_{j_1}\rangle|\phi^{3;\kappa_2}_{j_2}\rangle$. 
  The equations of motion (\ref{eqn_top},\ref{eqn_nontop}) can be simplified as follows 
  \begin{align}
 \label{eq_top2}
  i\partial_t A^1_I &= \sum_\kappa\sum_j \big[\sum_{p,q}(H^{3;\kappa}_0)_{pq}(\tilde Q^{\kappa}_{pq})_{i_\kappa j}+\frac{1}{2}\sum_{p_1,p_2,q_1,q_2}(V^{3;\kappa})_{p_1p_2q_1q_2}(\tilde P^{\kappa}_{p_1p_2q_1q_2})_{i_\kappa j}\big] A^1_{I^\kappa_j}\\ \nonumber
  &+\sum_{\kappa_1<\kappa_2}\sum_{j_1,j_2}\big[\sum_{p_1,p_2,q_1,q_2}(W^{3;\kappa_1,\kappa_2})_{p_1,p_2,q_1,q_2}(\tilde Q^{\kappa_1}_{p_1q_1})_{i_1j_1}(\tilde Q^{\kappa_2}_{p_2q_2})_{i_2j_2}\big]A^1_{I^{\kappa_1\kappa_2}_{j_1j_2}}.   
  \end{align}
  $I$ is again defined as an array $(i_1,i_2,...,i_S)$, while $I^\kappa_j$ and $I^{\kappa_1\kappa_2}_{j_1j_2}$
  are the arrays obtained by replacing $i_\kappa$ in $I$ by $j$ and $i_{\kappa_1}$, $i_{\kappa_2}$
  by $j_1$, $j_2$ in $I$, respectively.
  
  On the species layer, the Hamiltonian terms with nontrivial contribution to the rhs of equation (\ref{eqn_nontop})
  for $A^{2;\kappa}_{i;\vec{n}}$ are single particle Hamiltonian and intraspecies interaction terms of the $\kappa$
  species as well as the interspecies interaction terms between the $\kappa$ species and other species, while the
  remaining Hamiltonian terms do not contribute to the rhs due to the projection operator. The equations of motion
  on the species layer can be obtained by substituting the corresponding inverse density matrices and the mean-field
  operators $\langle \hat H\rangle^{2;\kappa}_{i,j}$ to equation (\ref{eqn_nontop}). The density matrices are calculated
  in the same way as in section \ref{ssubsec_eq}, and the mean-field operator $\langle \hat H\rangle^{2;\kappa}_{i,j}$,
  keeping only the nontrivial terms, is $\langle \hat H^0_\kappa+\hat V\rangle^{2;\kappa}_{i,j}+\langle \hat W_\kappa\rangle^{2;\kappa}_{i,j}$,
  and the mean-field operator for single particle 
Hamiltonian and 
intraspecies interaction terms are calculated as 
  \begin{align}
  \langle \hat H^0_{\kappa}+\hat V_{\kappa}\rangle^{2;\kappa}_{i,j}=(\rho^{2;\kappa})_{ij}*(\hat H^{3;\kappa}_0+\hat V^{3;\kappa}).
  \end{align}
   The mean-field operator for interspecies interaction is defined as
   $\langle \hat W_\kappa\rangle^{2;\kappa}_{i,j}\equiv\sum_{\nu\neq\kappa}\langle \hat W_\kappa\rangle^{2;\kappa,\nu}_{i,j}$, with
   \begin{align}
    \langle\hat{W}\rangle^{2;\kappa,\nu}_{ij} &= \sum_{\vec n|N_\kappa-1}\sum_{p,q=1}^{m_\kappa} Q_{\vec n}(p,q)|\vec n+\hat p\rangle_\kappa\langle \vec n+\hat q|
     \big[\sum_{r_\kappa=1}^{\mathscr M_\kappa}(A^{3;\kappa}_{p;r^\kappa})^*A^{3;\kappa}_{q;r^\kappa}(W^{\kappa,\nu}_{ij})(r^\kappa)\big],\\\nonumber
    (W^{\kappa,\nu}_{ij})(r^\kappa) &= \sum_{I^{\kappa\nu}}\sum_{r,s=1}^{M_\nu}(A^1_{I^{\kappa\nu}_{ir}})^*A^1_{I^{\kappa\nu}_{js}}\sum_{p,q=1}^{m_\mu}(\tilde Q^{\nu}_{pq})_{rs}\sum_{l_\nu=1}^{\mathscr M_\nu} (A^{3;\nu}_{p;l^\nu})^*A^{3;\nu}_{q;l^\nu}W_{\kappa,\nu}(r^\kappa,l^\nu)
  \end{align}
  
  The equations of motion for the particle layer are then obtained by substituting the corresponding
  inverse density matrix and mean-field operators to equation (\ref{eqn_nontop}). The mean-field operators
  for the single particle Hamiltonian, intraspecies and interspecies interactions are obtained as 
 \begin{align}
  \langle \hat H^0_{\kappa}\rangle^{3;\kappa}_{kj} &= \rho^{3;\kappa}_{ij}*\hat{H}^0_{\kappa},\\\nonumber
  \langle \hat V_{\kappa}\rangle^{3;\kappa}_{kj} &=  \frac{1}{N_\kappa}\sum_{r^\kappa}(V^\kappa_{kj})(r^\kappa)|r^\kappa\rangle\langle r^\kappa|\\\nonumber
  \langle \hat W_{\kappa,\nu}\rangle^{3;\kappa}_{kj}&= \frac{1}{N_\kappa}\sum_{r^\kappa}\big[\sum_{rs}(\tilde Q^\kappa_{kj})_{rs} (W^{\kappa,\nu}_{rs})(r^\kappa)\big]|r^\kappa\rangle\langle r^\kappa|.
  \end{align}
  $(V^\kappa_{kj})(r^\kappa)$ takes on the following appearance 
    \begin{align}
 (V^\kappa_{kj})(r^\kappa) &=\sum_{rs}(\rho^{2;\kappa})_{rs}\big\{\sum_{pq}(\tilde P^\kappa_{kpjq})_{rs}\big[\sum_l V_\kappa(r^\kappa,l)(A^{3;\kappa}_{p;l})^*A^{3;\kappa}_{q;l}\big]\big\}.\\\nonumber
    \end{align}

\bibliographystyle{unsrt}


\end{document}